\documentclass[sn-mathphys-num, twocolumn]{sn-jnl}
\usepackage{graphicx}%
\usepackage{multirow}%
\usepackage{amsmath,amssymb,amsfonts}%
\usepackage{amsthm}%
\usepackage{mathrsfs}%
\usepackage[title]{appendix}%
\usepackage{xcolor}%
\usepackage{textcomp}%
\usepackage{manyfoot}%
\usepackage{booktabs}%
\usepackage{algorithm}%
\usepackage{algorithmicx}%
\usepackage{algpseudocode}%
\usepackage{listings}%
\usepackage{braket}
\usepackage{caption}
\usepackage{makecell}
\usepackage{chemformula}
\usepackage{placeins}
\usepackage{bibunits}
\global\let\orcidlogo\relax
\usepackage{orcidlink}
\defaultbibliographystyle{sn-mathphys-num-custom}

\newcommand{\boldtheta}{{\boldsymbol\theta}}

\usepackage{comment}
\theoremstyle{thmstyleone}%
\theoremstyle{thmstyletwo}%

\theoremstyle{thmstylethree}%

\begin{document}
\title[]{Ab-initio simulation of excited-state potential energy surfaces with transferable deep quantum Monte Carlo}

\author[1]{\fnm{Zeno} \sur{Sch{\"a}tzle}\orcidlink{0000-0002-5345-6592}}
\equalcont{These authors contributed equally to this work.}

\author[1]{\fnm{P.} \fnm{Bern{\'a}t} \sur{Szab{\'o}}\orcidlink{0000-0003-1824-8322}}
\equalcont{These authors contributed equally to this work.}

\author[1]{\fnm{Alice} \sur{Cuzzocrea}\orcidlink{0000-0001-7446-9643}}

\author[1]{\fnm{Mat\v{e}j} \sur{Mezera}\orcidlink{0009-0003-0047-488X}}

\author*[1,2,3,4]{\fnm{Frank} \sur{Noé}\orcidlink{0000-0003-4169-9324}}\email{frank.noe@fu-berlin.de}

\affil[1]{\orgdiv{Department of Mathematics and Computer
Science}, \orgname{Freie Universität Berlin}, \orgaddress{\city{Berlin}, \country{Germany}}}

\affil[2]{\orgdiv{Department of Physics}, \orgname{Freie Universität Berlin}, \orgaddress{\city{Berlin}, \country{Germany}}}

\affil[3]{\orgname{Microsoft Research, AI for Science}, \orgaddress{\city{Berlin}, \country{Germany}}}

\affil[4]{\orgdiv{Department of Chemistry}, \orgname{Rice University}, 
\orgaddress{\city{Houston}, \state{Texas}, \country{USA}}}

\abstract{The accurate quantum chemical calculation of excited states is a challenging task, often requiring computationally demanding methods. When entire ground and excited potential energy surfaces (PESs) are desired, for instance to predict the interaction of light excitation and structural changes, one is often forced to use cheaper computational methods at the cost of reduced accuracy. Here we introduce a method for the geometrically transferable optimization of neural network wave functions that leverages weight sharing and dynamical ordering of electronic states.
Our method enables the efficient prediction of ground and excited-state PESs and their intersections at the highest accuracy, demonstrating up to two orders of magnitude cost reduction compared to single-point calculations.
We validate our approach on four challenging excited-state PESs, namely ethylene, the carbon dimer, the methylenimmonium cation, and a rubredoxin active site model containing 96 electrons, illustrating the potential of transferable deep-learning QMC as a practical framework for studying electronic excitations in molecules.
}

\keywords{Excited States, Quantum Monte Carlo, Deep Learning, Parameter Sharing}

\maketitle
\begin{bibunit}
\begin{figure*}[!t]
    \centering
    \includegraphics[width=0.85\textwidth]{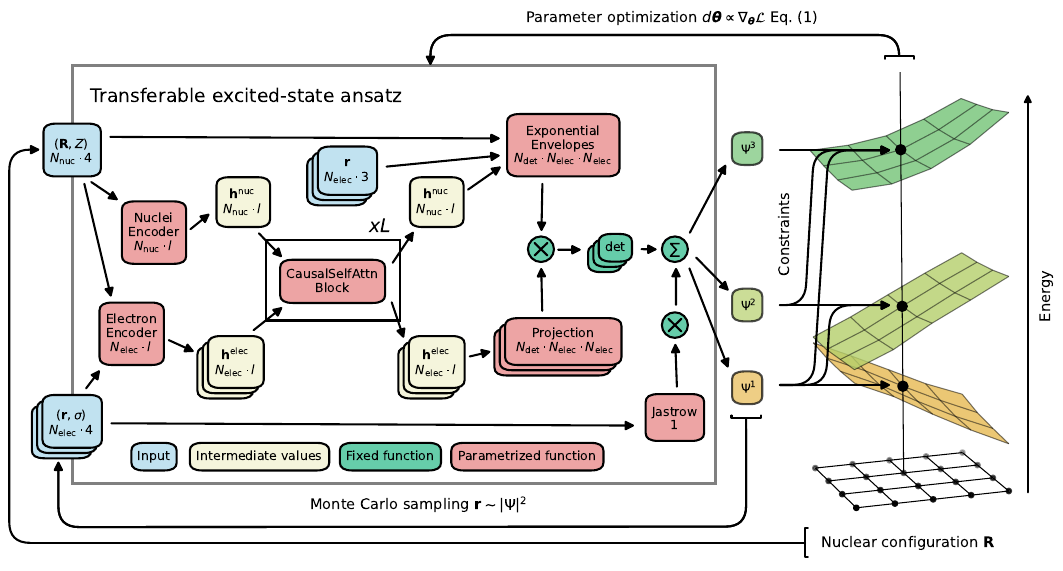}
    \caption{\textbf{Neural network architecture for the transferable excited-state wave function with orthogonality constraints.} Diagram of the transferable wave function. Nuclear positions inform the electron representations in the \textsc{CausalSelfAttn} block, introducing an explicit dependency on the molecular configuration. If parameters are shared between states, the same \textsc{Electron Encoder}, \textsc{Nuclei Encoder} and  \textsc{CausalSelfAttn} block are applied for each batch of electron coordinates, respectively. The orthogonality of the states is implemented by ordering wave functions based on energy and applying a directional overlap penalty \eqref{eqn:loss_function}.
    }
    \label{fig:architecute}
\end{figure*}

Photochemistry plays a fundamental role in biological systems, driving essential processes such as vision, photosynthesis, and molecular photoprotection \cite{gozem2017,curutchet2017, westermayr2022}. Light-driven phenomena are also key to technological advancements, ranging from material design and chemical processing \cite{risko2011,shaw2016}  to biomedical technologies such as molecular motors and photo-controlled drug delivery \cite{deng2024,alvarez-lorenzo2009}.

Despite the critical importance of these processes, their theoretical study is hindered by the need for accurate \textit{ab-initio} descriptions of electronic excited states.
Most quantum chemistry methods have been developed for the calculation of electronic ground states and their extensions to excited states are either limited or highly expensive and often require expert knowledge \cite{serrano-andres2005, gonzalez2021}.
These problems are further amplified when targeting a coherent description of excited state potential energy surfaces (PESs), which are the key ingredient to simulating the photodynamics of molecules \cite{westermayr2021}. 
Time-dependent density functional theory
is widely used for excited-state characterization but struggles with delocalized excitations, conical intersections, high-energy states, multi-electron processes, and charge transfer \cite{casida2012, herbert2023}. Similarly, excited-state extensions of coupled cluster methods, which rely on a single reference determinant, become unreliable in systems with strong multi-reference character \cite{kohn2007}.
On the other hand, most multi-reference methods require the definition of an active space, a process that often relies on chemical intuition to select relevant orbitals and becomes increasingly challenging when considering structural flexibility, for instance in dynamical simulations \cite{stein2016}.
Even if sophisticated automatic tools are used to obtain appropriate active spaces \cite{bensberg2023,stein2016}, accurately describing dynamic correlation often necessitates the use of multi-reference coupled cluster or (semi-stochastic) configuration interaction (CI) methods.
However, these approaches remain less established than their single-reference counterparts and exhibit steep computational scaling with system size \cite{lischka2018}.

Variational quantum Monte Carlo (VMC) offers a promising alternative to traditional methods, given its black-box nature, favorable scaling with system size, and flexibility to address virtually any electronic structure problem \cite{foulkes2001,lester2009,feldt2021}.
Neural network based VMC has recently been shown to yield high accuracy in challenging excited-state problems \cite{entwistle2023,szabo2024, pfau2020}, including conical intersections and double excitations.
Despite this success, the simulation of excited-state PESs remains costly due to the need of performing large numbers of independent single-point calculations.
At the same time, sharing neural network wave function parameters across molecular geometries (geometric transferability) \cite{scherbela2022,gao2022} was shown to be both accurate and computationally efficient in simulating ground state PESs and proved beneficial for generalizing across molecules \cite{scherbela2024,gao2023a, gao2024, foster2025} as well as spin lattice Hamiltonians \cite{rende2025a}.
Extending geometric transferability of neural network wave functions to excited state calculations promises a wide range of benefits, including reduced computational costs, improved error cancellation, as well as more stable optimization processes.
Such an approach could provide unparalleled \textit{ab-initio} access to all electronic properties of even the most challenging excited states at a systematically refinable accuracy.

In this work, a method for the transferable VMC optimization of neural network wave functions for excited-state PESs is presented.
Exploiting similarities of wave functions across molecular geometries, the efficiency of computing excited-state PESs is increased substantially over single-point simulations, with a speedup of close to two orders of magnitude being demonstrated.
Further improvements are attained by introducing parameter sharing between electronic states and a dynamic state ordering mechanism for the orthogonalization of states based on the per-geometry energy.
The method yields a coherent \textit{ab-initio} description of excited-state PESs with full access to electronic properties, which we demonstrate on four challenging applications -- the ground and singlet excited states of the ethylene torsion and pyramidalization PESs, the eight lowest-lying states of the carbon dimer dissociation, the two-dimensional excited state PESs of the three lowest singlet states of the methylenimmonium cation, and the spin-state energetics of a 96-electron model of the rubredoxin active site.

\section*{Results}\label{sec2}

\subsection*{Transferable ab-initio optimization of excited states}

In traditional quantum chemistry, calculating excited-state PESs involves the costly simulation of multiple wave functions for a large number of molecular geometries.
Taking advantage of similarities between such wave functions and leveraging the high expressivity of neural network ansatzes, we introduce a method that generalizes across molecular configurations as well as electronic excited states, achieving a balanced description of excited-state PESs at a much reduced cost.

The framework combines multiple components, which we condensed in a loss function for variational Monte Carlo optimization:
\begin{strip}
\begin{equation}
    \mathcal{L}(\boldtheta_1,...,\boldtheta_{N_\text{s}}\underbrace{
        \vphantom{\sum_{\mathbf{R} \in \boldsymbol{\Re}}}
        ,\boldtheta^\text{s})
    }_{\substack{\text{shared} \\ \text{params}}} =
    \underbrace{
        \sum_{\mathbf{R} \in \boldsymbol{\Re}}
    }_{\substack{\text{geometry} \\ \text{transfer}}} \,
    \underbrace{
        \vphantom{\sum_{\mathbf{R} \in \boldsymbol{\Re}}}
        \sum_{i}^{N_\text{s}}
    }_{\substack{\text{electronic} \\ \text{states}}} 
    \Big[
        \underbrace{
            \vphantom{\sum_{\mathbf{R} \in \boldsymbol{\Re}}}
            \braket{\hat{H}}_\mathbf{R}^i
        }_{\substack{\text{energy} \\ \text{expectation}}} +
        \underbrace{
            \vphantom{\sum_{\mathbf{R} \in \boldsymbol{\Re}}}
            \beta \braket{\hat{S}^2}_\mathbf{R}^i
        }_{\substack{\text{spin} \\ \text{penalty}}} +
        \underbrace{
            \vphantom{\sum_{\mathbf{R} \in \boldsymbol{\Re}}}
            \sum_{j}^{E^i_{\bf R} > E^j_{\bf R}} \alpha_{ij} |\braket{\Psi^{i} | \bar \Psi^{j}}_\mathbf{R}|^2
        }_{\substack{\text{dynamic} \\ \text{overlap penalty}}}
    \Big]\;.
    \label{eqn:loss_function}
\end{equation}
\end{strip}
The loss comprises the molecular energy expectation value $\braket{\hat{H}}$, and the spin-penalty  $\braket{\hat{S}^2}$ \cite{szabo2024,penalty} for each electronic state $i$ and configuration $\bold R$. 
The parameters $\boldtheta_i$, one for each of the $N_\text{s}$ states, are optimized simultaneously across a dataset of molecular geometries $\boldsymbol{\Re}$, constituting a substantial advancement over the conventional independent treatment \cite{entwistle2023,pfau2024,szabo2024} in terms of convergence speed and accuracy.
We will refer to the optimization of geometry dependent electronic wave functions across a dataset of molecular configurations as \textit{transferable training} throughout the manuscript.
Additionally, partial sharing of parameters $\boldsymbol{\theta}^\text{s}$ between different electronic states leads to enhanced error cancellation (see Sec.~\ref{sec:methods:parameter_sharing}), improving convergence and increasing the accuracy in excitation energies.
This sharing of model parameters between the wave functions of different electronic states is referred to as \textit{parameter sharing}.

While previous penalty-based QMC algorithms enabled the computation of multiple excited states via an overlap penalty term between states with a fixed order,
the last term of \eqref{eqn:loss_function} orders states dynamically based on the per-geometry energy $E_{\bf R}$ (further details in Sec.~\ref{sec:methods:reordering}).
This enhances the robustness of our method in the presence of level crossings and conical intersections -- crucial albeit challenging points for quantum chemistry -- promoting the continuity of the ansatzes at points where the energy ordering of the states changes.
Notably, due to the regularization properties and improved convergence of the transferable training, the simulation of excited state PESs is further simplified by eliminating the need for supervised pretraining on external quantum chemistry methods.
Additionally, incorporating an explicit parametric dependence of the ansatz on nuclear positions further increases expressivity and promotes stable convergence.
Concretely, the self-attention block $\text{SelfAttn}({\bf h}^\text{elec})$ of the Psiformer \cite{glehn2023} is extended by considering both nuclear and electronic nodes
\begin{equation}
    \text{CausalSelfAttn}(\text{Concat}({\bf h}^{\text{nuc}}, {\bf h}^{\text{elec}})) \, ,
\end{equation}
with causal attention preventing information from flowing from electrons to nuclei.
After applying $L$ layers of self-attention, the final electron representations ${\bf h}^{\text{elec}}$  are projected to give many-body orbitals and nuclei representations ${\bf h}^{\text{nuc}}$ are used to predict parameters of the orbital envelopes.
The architecture of the ansatz is depicted on Fig.~\ref{fig:architecute} and further discussed in Sec. \ref{sec:ansatz}.

In the following, we demonstrate the benefits of these methodological advancements on several challenging excited-state PESs. Our results are compared to state-of-the-art multi-reference quantum chemistry methods and single-point deep-learning QMC simulations.
As a quantitative measure of accuracy, we propose the mean absolute error (MAE) of relative energies, computed between target and reference results after shifting the latter such that their mean energies across all geometries and all states are aligned.

\subsection*{Unified description of excited state processes}

\begin{figure*}[t]
    \centering
    \includegraphics[width=\textwidth]{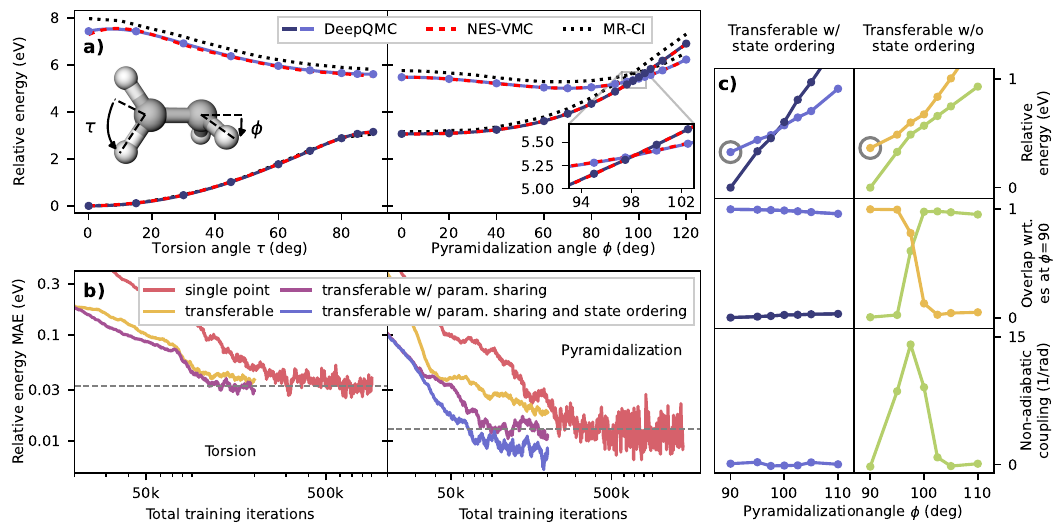}
    \caption{\textbf{Lowest-lying singlet PESs for ethylene relaxation.} 
    Simulation of the ground and first singlet excited state of ethylene along the torsion around the C--C bond and pyramidalization of a CH$_2$ group.
    \textbf{a)}: Relative energy with respect to the equilibrium geometry. DeepQMC results for transferable optimization across molecular geometries with parameter sharing across electronic states and dynamic state ordering are presented alongside highly accurate single-point NES-VMC simulations \cite{pfau2024} and MR-CI results \cite{barbatti2004}. Sampling errors are smaller than the marker size.
    \textbf{b)}: Convergence of the DeepQMC MAEs to NES-VMC results in relative energy. Different variants of transferable DeepQMC optimization are compared to single-point calculations performed in earlier work \cite{szabo2024}. 
    The horizontal axis indicates the total cumulative iterations across all geometries of the torsion and pyramidalization PES, respectively.
    The dashed horizontal line gives the accuracy of evaluating the final single-point wave functions.
    \textbf{c)}: Relative energies with respect to $\phi=90^\circ$ around the conical intersection (top panels), intermolecular overlaps (Eq.\ref{eq:intermolecular_overlap}, center panels) and non-adiabatic couplings (bottom panels) for transferable DeepQMC runs with parameter sharing across electronic states and dynamic state ordering and without parameter sharing across electronic states and fixed ordering.
    }
    \label{fig:ethylene}
\end{figure*}

Ethylene serves as a minimal model system for photoisomerization processes and internal energy conversion and its excited-state PES has been studied extensively \cite{ben-nun2000, barbatti2004}.
After photo-excitation to the first singlet excited state, ethylene undergoes a torsion around the carbon-carbon bond and a pyramidalization of one of the CH$_2$ units, leading to a conical intersection from which the system relaxes back to the ground state equilibrium geometry.
Computationally modeling the non-radiative relaxation pathway is challenging due to strong electronic correlation in the excited state, increasing multi-reference character upon approaching the conical intersection, and the existence of intersystem crossings.
It has been shown that single-reference methods such as SR-CI qualitatively fail at describing several features of the ethylene PES \cite{shao2003, barbatti2004, ben-nun2000}.
Deep-learning QMC, on the other hand, does not suffer from these limitations \cite{hermann2020, pfau2020} and previous, single-point applications of excited state deep-learning QMC methods have modeled the PESs of ethylene with high accuracy \cite{entwistle2023,szabo2024,pfau2024}.

In this work, two transferable deep-learning QMC optimizations are run for the 1D PESs along the torsion (9 geometries) and pyramidalization (14 geometries) of ethylene \cite{barbatti2004}.
The spin penalty \eqref{eqn:loss_function} is utilized to target singlet states only.
The results of the transferable optimization with parameter sharing and dynamic state ordering displayed in Fig.~\ref{fig:ethylene}, are in excellent agreement with the single-point natural excited state (NES) QMC reference \cite{pfau2024}. 
After a total of 200k training iterations evenly split across all geometries, we obtain a MAE for the relative energies of 29.4(1)~meV along the torsion PES, and 4.7(7)~meV along the pyramidalization PES.
This improves over previously obtained, single-point, penalty-based DeepQMC results \cite{szabo2024} of 33.8(3)~meV and 13.1(3)~meV, while using about five times less computational resources and without employing supervised pretraining with respect to CASSCF baselines.
The location of the conical intersection of the pyramidalization PES ($98.5^\circ$) is excellently reproduced, while due to the dynamic ordering of the states the ansatzes are continuous across the conical intersection.

The beneficial effects of dynamic state ordering and parameter sharing are highlighted in Fig.~\ref{fig:ethylene}b, comparing the convergence of the relative energy MAEs during optimization with and without parameter sharing and dynamic state ordering.
First, dynamic state ordering facilitates convergence in the presence of a conical intersection, reaching lower errors faster, than other transferable calculations on the pyramidalization PES.
Second, sharing the parameters of the main electron-nucleus transformer across the two ansatzes leads to improved convergence of the MAE of relative energies compared to independent transferable wave functions on both PESs.
The effect of the dynamic state ordering is furthermore illustrated in Fig.~\ref{fig:ethylene}c, where energies, intermolecular overlaps and non-adiabatic couplings at the conical intersection are shown.
While the runs with the option turned on are continuous through the crossing and intermolecular overlaps stay close to one within a state and close to zero with respect to the second state, runs without dynamic state ordering display a discontinuity at the conical intersections, where the intermolecular overlaps indicate a flip of the states.
This effect also shows in the computation of non-adiabatic couplings (NACs) relevant to surface-hopping simulations. 
NACs can be obtained by approximating the overlap gradient along the pyramidalization angle $\phi$ using finite differences
$h_{ij}(\phi) = \nabla_{\phi'} \langle \psi_i(\phi) |  \psi_j(\phi')\rangle|_{\phi =\phi'}\approx\frac{1}{2 \Delta \phi} (\langle \psi_i(\phi) |\psi_j (\phi + \Delta \phi)\rangle - \langle \psi_i(\phi) |\psi_j (\phi - \Delta \phi)\rangle)$ \cite{plasser2016} (here we used non-uniform finite difference).
We find that transferable simulations employing dynamic state ordering yield quasi-diabatic representations. These are electronic states that are continuous along the one-dimensional cut through the conical intersection. Consequently, the NAC remains small at the conical intersection. By contrast, simulations executed with a fixed ordering exhibit pronounced coupling at the conical intersection, due to state discontinuities.

\subsection*{Efficiently simulating challenging electronic structures}

\begin{figure*}[t]
    \centering
    \includegraphics[width=\linewidth]{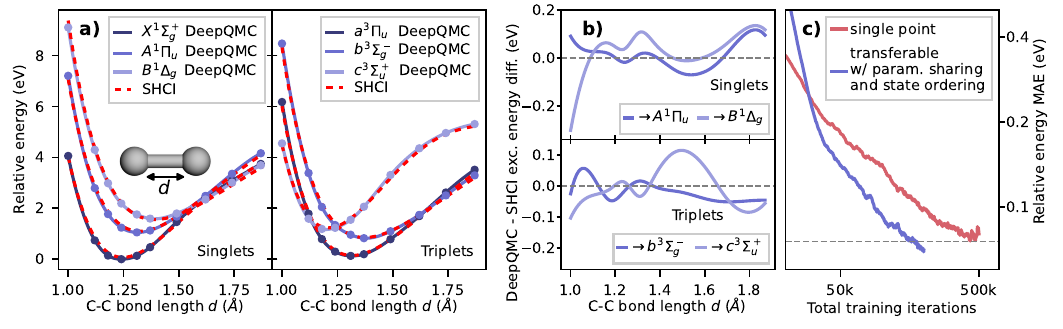}
    \caption{
    \textbf{Carbon dimer dissociation curves along the lowest-lying eight electronic states.}
    Transferable DeepQMC ansatzes were trained jointly across ten geometries, with bond lengths ranging from 1.0 to 1.9 \AA and sharing parameters across electronic states.
    Semistochastic heat-bath configuration interaction calculations in quintuple-$\zeta$ basis, extrapolated to the full configuration interaction limit are used as reference \cite{holmes2017}.
    \textbf{a)}: The final relative energies obtained with transferable DeepQMC, for the lowest-lying four singlet and four triplet states of the carbon dimer.
    Both DeepQMC and reference curves were smoothed by cubic interpolation.
    \textbf{b)}: Differences between DeepQMC and semistochastic heat-bath configuration interaction in the vertical excitation energies from the ground state of the respective spin sector.
    \textbf{c)}: Convergence of the mean absolute error of relative energies between the eight lowest-lying states of the carbon dimer.
    Geometrically transferable optimization is compared with single-point training performed in earlier work \cite{szabo2024}.
    The horizontal axis indicates the total cumulative iterations across all geometries. The dashed horizontal line marks the evaluated final accuracy of the single-point simulations.
    }
    \label{fig:carbon-dimer}
\end{figure*}

Computing the numerous low-lying electronic states of the dissociation of the carbon dimer constitutes a formidable challenge to quantum chemistry methods \cite{,wouters2014,holmes2017}.
The difficulty lies in the simultaneous presence of strong static and dynamic correlation \cite{mazin2021}, giving rise to a number of exotic electronic properties such as charge-transfer type bonding \cite{su2011}, and the presence of low-lying doubly excited states \cite{bruna1991}.

The benefits of optimizing excited state deep-learning QMC ansatzes jointly across molecular geometries are showcased by modeling the lowest-lying eight states of the carbon dimer simultaneously for ten nuclear separations ranging from 1.0 to 1.9 \AA.
Utilizing the spin penalty term in the loss function \eqref{eqn:loss_function}, the four singlet and four triplet states can be treated separately.
The parameters of the main electron-nucleus transformer are shared across the four electronic states in each simulation.
The resulting PESs are shown on Fig.~\ref{fig:carbon-dimer}a, along with reference values obtained with the semi-stochastic heat bath configuration interaction (SHCI) \cite{holmes2017} method.
It is clear that joint DeepQMC optimization is capable of modeling all eight states, along the entire range of bond lengths.
The convergence of the MAE of relative energies during the training process is plotted on Fig.~\ref{fig:carbon-dimer}c, for single-point \cite{szabo2024}, and transferable DeepQMC optimizations.
The final MAEs of the two setups are similar, being 75(2) meV for single-point, and 69(1) meV for transferable optimization.
However, joint training offers large cost savings, converging to this accuracy substantially faster, reaching the final MAE of the single-point calculation in about three times fewer iterations.
This is all the more remarkable as unlike the single-point run, transferable optimization was not preceded by supervised pretraining.
It's worth emphasizing that the above results pertain to the accuracy of relative energies, as opposed to the absolute energy of each point along the PESs.
While the former is vastly more important in practice, we note that the  model trained jointly for 200k iterations yields on average 170(2) meV higher absolute energies than the single-point one.
The larger absolute energy error for the transferably trained ansatz is explained by the fact that it has (roughly) the same capacity as the single-point ansatz, but has to simultaneously represent a range of geometries, which necessarily leaves less expressivity for each geometry.
On the other hand, this highlights the improved error cancellation properties afforded by joint optimization, which arrives at lower relative energy errors than single-point training.
At the same time, the absolute energies of the transferable ansatz were still converging when the optimization was stopped, indicating the possibility of further reducing the total energy error with extended optimization.

\begin{figure*}[!t]
\centering
\includegraphics[width=\linewidth]{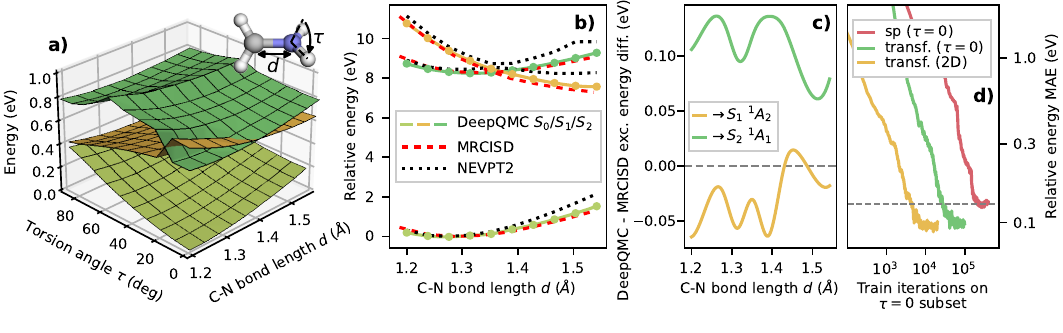}
\vspace{-8pt}
\caption{\textbf{2D singlet PESs of the methylenimmonium cation.}
Transferable DeepQMC simulations on \ch{CH2NH2+} for varying \ch{C-N} bond lengths (1.2--1.54\,\AA{}) and torsion angles $\tau$ (0--90$^\circ$). 
\textbf{a)}: Relative energies of the three lowest-lying singlet PESs of the  \ch{CH2NH2+} on a uniform 10$\times$10 grid.
\textbf{b)}: 1D slice of the 2D PES at $\tau=0$, comparing transferable DeepQMC results with MR-CISD+Q/SA-9-CAS(6,4)@d-aug-cc-pVDZ data from Ref.~\cite{barbatti2006} and NEVPT2/CAS(10,10))@cc-pvQZ simulations obtained with PySCF \cite{sun2020}.
\textbf{c)}: Difference between the DeepQMC and MRCISD results on the $\tau=0^\circ$ slice in vertical excitation energies from the ground state.
\textbf{d)}: Convergence of the mean absolute error of relative energies between the three lowest-lying singlet states evaluated at ten configurations with $\tau=0$ (\ch{C-N} bond length between 1.2--1.54\,\AA), using MR-CISD~\cite{barbatti2006} as a reference. 
Single-point and two transferable neural network VMC optimizations are compared.
The first transferable training was carried out on the 10$\times$10 grid (varying \ch{C-N} and $\tau$), while the second only on the 10-point subset at $\tau=0$. The horizontal axis indicates the total cumulative iterations across all geometries considered, re-normalized to take into account differences in the electron batch sizes. The dashed horizontal line marks the final accuracy of the single-point simulations.
}
\label{fig:ch2nh2p}
\end{figure*}

Even with dynamic state ordering and sharing of network parameters across states, we did not succeed in keeping the continuity of the $X^1\Sigma^+_g$ and $A^1\Pi_u$ states at large bond lengths, where they are very nearly degenerate for a large section of the PES ($>1.6$ \AA).
Increasing the value of the overlap penalty scaling factor $\tilde \alpha$ did not recover the correct energy ordering, while very large values could even degrade overall performance \cite{szabo2024}.
The effect of increasing $\tilde \alpha$ is diminished on these states by the very form of the overlap scaling factor (see Section \ref{sec:methods:reordering}), which was designed with conical intersections in mind, where the energies of crossing states approach each other rapidly and are nearly-degenerate only for a narrow range of geometries.
In contrast, the curves of the $X^1\Sigma^+_g$ and $A^1\Pi_u$ C\textsubscript{2} states run nearly parallel, resulting in low overlap penalty scales for such a wide range of bond lengths, that our flexible neural network ansatzes can change character without incurring substantial penalties in the loss function.
While finding an improved form for the overlap scaling factor certainly warrants further research, state crossings of the first kind where the current form performs excellently are by far the more common, therefore our proposed methodology is already widely applicable.
Furthermore, the parameter sharing and dynamic state ordering features do have a beneficial effect on the continuity of the $a^3\Pi_u$ and $c^3\Sigma^+_u$ states at their crossing, where continuity is preserved consistently only with these settings active.

\subsection*{Higher computational efficiency gains for multidimensional potential energy surfaces}

We continue by presenting results on the excited states of the methylenimmonium cation, \ch{CH2NH2+},
the simplest model of protonated Schiff bases undergoing isomerization.
As such, it is extensively studied in the context of photodynamics \cite{barbatti2006,li2017}, and more recently became the target of neural network force fields for excited states \cite{westermayr2019}.
Following a $\pi \rightarrow \pi^*$ excitation, the system can first relax to the first excited state via elongation of the \ch{C-N} bond combined with slight bi-pyramidalization, and then to the ground state via torsional motion.
Accurately capturing this relaxation remains a challenge for CASSCF,
overshooting the $\pi\pi^*$ state in the Franck--Condon region~\cite{barbatti2006}.
Accordingly, the more computationally expensive MR-CISD calculations are to date the 
\textit{de facto} references for the study of this system~\cite{barbatti2006}.

In contrast to the small scale study of ethylene and carbon dimer, we now turn to a two-dimensional problem, covering key molecular configurations relevant for the relaxation of \ch{CH2NH2+}.
We generate a $10 \times 10$ grid of nuclear configurations, spanning the two most relevant degrees of freedom in the problem, the torsion angle $\tau$ and the \ch{C-N} bond length. 
Unlike in previous examples, here we employ ccECP-type pseudopotentials~\cite{bennett2017, schatzle2023}. 
While not strictly necessary, this choice speeds up convergence and reduces the memory required for storing electron position samples for each molecular configuration. 
We found that sharing network parameters across states did not provide a clear advantage. Therefore we maintain separate parameters for each state. We believe that the use of pseudopotentials diminishes the benefits of parameter sharing, as a substantial portion of the shared wave function features is lost when replacing core electrons with effective potentials.

\begin{figure*}[t]
    \centering
    \includegraphics[width=\linewidth]{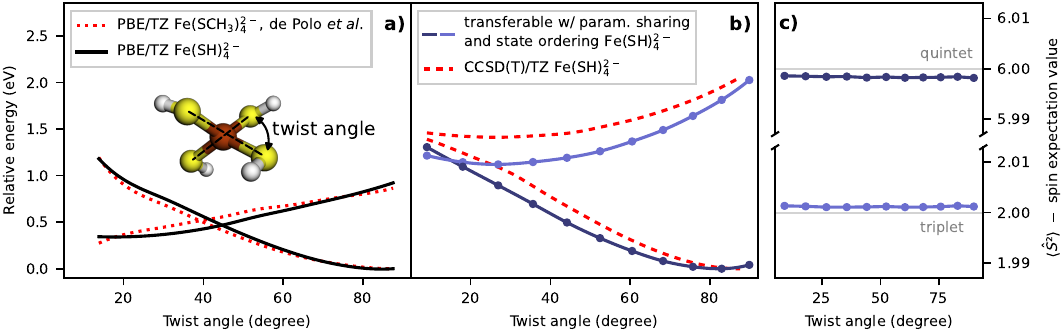}
    \caption{
        {\bf Crossing of the lowest lying quintet and triplet states of Fe(SH)$_4^{2-}$}.
        The molecular geometries are defined by the twist angle running from 9.1$^\circ$ (nearly planar) to $89.9^\circ$ (almost tetrahedral).
        \textbf{a)}: Validation of the use of the smaller Fe(SH)$^{2-}_4$ model compared to the original Fe(SCH$_3$)$^{2-}_4$ model by comparing PBE relative spin-state energies in the def2-TZVP basis \cite{depolo2016}.
        \textbf{b)}: Qualitative comparison of DeepQMC and reference CCSD(T)@def2-TZVP relative energies for the lowest-lying triplet and quintet states of the Fe(SH)$^{2-}_4$ model.
        DeepQMC calculations are carried out transferably across all eleven geometries, sharing parameters across electronic states and utilizing overlap and spin penalties (to target the triplet state). An extended discussion of the reliability of the reference simulations is presented in Supplementary Section \ref{sec:app:rubredoxin}.
        \textbf{c)}: Spin expectation values of the DeepQMC wave functions. The spin contamination of about 0.001-0.002 can be considered negligible.
    }
    \label{fig:rubredoxin}
\end{figure*}

In Fig.~\ref{fig:ch2nh2p}a, we present the 2D PES of the methylenimmonium cation on a uniform grid, the \ch{C-N} bond length ranging from 1.2 to 1.54 \AA, with torsion angles between $0^\circ$and $90^\circ$.
First, the qualitative agreement across the whole grid with the MR-CISD+Q/SA-9-CAS(6,4)@d-aug-cc-pVDZ reference results from Ref.~\cite{barbatti2006} is noted. Our model correctly captures the approach of the ground and first excited states with increasing torsion angle, as well as the crossing of the first and second excited-state PESs at $\tau=0^\circ$. In Fig.~\ref{fig:ch2nh2p}b we further inspect a slice of the grid at $\tau=0^\circ$ and directly compare to the MR-CISD reference.
Notably, dynamic state ordering enables each ansatz to follow the respective state's character. Our results are in good agreement with Ref.~\cite{barbatti2006}, as the positions of the conical intersections coincide, while the remaining small differences are comparable to the accuracy of the reference method and equally plausibly may reflect its own limitations.

To assess the cost savings achieved by training jointly across molecular geometries, we compare the convergence of single-point DeepQMC calculations with two transferable models in Fig.~\ref{fig:ch2nh2p}d, one trained on the 100-point grid described above, the other trained only on the subset of ten molecular configurations with $\tau=0^\circ$. Similar to Fig.~\ref{fig:ethylene}b and \ref{fig:carbon-dimer}b, we utilize the MAE of relative energies, computed for the ten geometries with $\tau=0^\circ$, to compare the convergence of the various DeepQMC calculations.
Unlike in the above studies, here the baseline single-point optimizations too were performed without supervised pretraining, facilitating a direct comparison. Once again we observe a substantial convergence speedup, by roughly one order of magnitude, when going from the single-point calculations to the transferable setup with ten molecules.  
An additional order-of-magnitude speedup is obtained when training the transferable wave function on the larger dataset, containing 100 molecular configurations.
We emphasize that the 10-point grid is a subset of the 100-point grid for a fixed torsion angle, and thus the 100-point grid covers a substantially larger section of the full PES.
Finally, we highlight that transitioning to a transferable training setup not only accelerates convergence, but also improves accuracy. This is evident from the evaluated MAE (with respect to MR-CISD) dropping from the 134(2) meV of the single-point calculation to 96(2) and 97(5) meV for the 10- and 100-point transferable wave functions, respectively.
Interestingly, the close agreement between the latter two suggests the robustness of our approach and raises the possibility that the discrepancy with the reference originates from its own limitations.
When simulating larger parts of molecular configuration space another increasingly relevant factor is the potential for convergence to incorrect states, either due to sections that are more prone to these difficulties or simply the increasing likelihood of a few out of many single-point simulations exhibiting erroneous convergence.
Here, transferable training can act as a regularizer. 
By correctly identifying the states for a subset of molecular configurations, the entire transferable model is driven towards proper convergence (more details can be found in Supplementary Section \ref{sec:app:methylenimmonium}).

\subsection*{Scalability to transition metal complexes with pseudohamiltonians}

It has been suggested that spin-forbidden transitions might play an important role in the biological function of the rubredoxin electron-transfer protein \cite{depolo2016}.
dePolo and coworkers \cite{depolo2016} used the PBE density functional to identify a crossing of the low-lying triplet and quintet states of the Fe(SCH$_3$)$_4^{2-}$ model of the rubredoxin active site, as it changes from planar to tetrahedral geometry, where non-negligible spin-forbidden transition rates were obtained via non-adiabatic transition state theory.
Here, we employ the slightly smaller Fe(SH)$_4^{2-}$ model with 96 electrons and build a one-dimensional potential energy surface analogous to the one originally investigated.
First, on Fig.~\ref{fig:rubredoxin}a) it is verified that the PBE functional identifies the same crossing of the triplet and quintet states for this smaller model as for the original Fe(SCH$_3$)$_4^{2-}$, thus validating the removal of the methyl groups.
Then, we employ our proposed methodology to describe the triplet and quintet states of the Fe(SH)$_4^{2-}$ model using deep-learning QMC.
The fast-moving core electrons of the iron and sulfur atoms are replaced with local pseudohamiltonians \cite{ichibha2023,fu2026}, and the spin-penalty is applied to one of the states to restrict it to the triplet spin sector.
A single, two-state DeepQMC simulation is performed, where the parameters of the electron-nucleus transformer are shared between the two states, and the spin-penalty is applied to the triplet state only.
As can be seen on Fig.~\ref{fig:rubredoxin}b), DeepQMC converges to PESs that are qualitatively different from those obtained with PBE, and are more similar to CCSD(T) ones, predicting a quintet ground state for all but the most planar geometries under investigation.
Indeed, it is known that generalized gradient functionals such as PBE tend to overly favor low-spin states \cite{bowman2012,swart2008}.
These results indicate that the molecular geometry of the triplet-quintet crossing in the twice negatively charged rubredoxin active site might be substantially different from what was previously assumed, with potential implications on the rate of the associated spin-forbidden transitions.
Lastly, Fig.~\ref{fig:rubredoxin}c) shows that the spin contamination of the DeepQMC wave functions is negligible (0.001-0.002), indicating that pure and continuous states are retained across the entire PES.
With Fe(SH)$_4^{2-}$ being the largest system for which transferable training and parameter sharing has been applied, this result underscores the potential in exploiting similarities across wave functions of different molecular geometries and electronic states via machine learning approaches.

\section*{Discussion}\label{sec12}
We have developed a computational method to simultaneously optimize the electronic wave functions and energies for ground and excited states of molecules with a range of nuclear configurations, thus obtaining \textit{ab-initio} ground- and excited-state PESs. 
This is achieved by designing a neural network wave function that is trained with VMC across the desired nuclear configurations and electronic states.
Due to the simultaneous representation of multiple geometries and states that is enabled by the expressivity of neural network wave functions, we achieve substantial reductions of computational cost when optimizing entire PESs compared to analogous methods where each single-point calculation is done from scratch. At the same time, our approach maintains extremely high accuracy, as shown in comparison with different bespoke quantum chemistry methods that have been carefully selected for each considered example.
The improved efficiency of joint training is demonstrated by estimating the challenging PESs of the ethylene relaxation, the carbon dimer, and the methylenimmonium cation using five, three and up to about a hundred times less compute, respectively, than the equivalent single-point deep-learning QMC calculations.
In addition to decreased computational costs, the transferable training improves the accuracy of relative energies, as evidenced on these three test systems.
The higher accuracy and stability are in large part due to the enhanced error cancellation which arises naturally if the same ansatz is used to describe all geometries in question.
Lastly, the scalability of the proposed method to larger systems is demonstrated by modeling two spin states of a 96-electron model of the rubredoxin active site.

An important aspect of geometrically transferable deep-learning QMC is the presence of state crossings, where the static ordering of states in the loss function can lead to a quasi-discontinuity in the state represented by a given network.
To promote the continuity of states and improve convergence, dynamic state ordering is introduced, which constructs the most appropriate loss function independently for each nuclear configuration.
Furthermore, the concept of sharing large parts of the neural network parameters between models of different electronic states is introduced.
If there are similarities between the different states, the shared parameters effectively receive more gradient information during training while correlation between the electron representations improves error cancellation, contributing to the improvement over single-point results on ethylene and the carbon dimer.

Despite these improvements, several challenges remain in simulating excited-state PESs with deep-learning QMC. 
First, our methodology targets fixed datasets of molecular geometries and, in its current form, scales only to a few hundred to a few thousand geometries. 
Extending it to high-dimensional PESs will require new Monte Carlo sampling schemes that combine electron and nuclei moves, together with a corresponding replacement of the exponential moving averages of the energies in the dynamic state reordering, for instance with a surrogate model. 
Second, simulating larger regions of configuration space and larger molecules risks state swapping with the highest-energy excited state at the boundary. 
While dynamic state reordering accounts for reordering within the subspace of the $n$ lowest-energy states, excited states entering from outside this subspace can still cause discontinuities in the wavefunction, reducing the efficiency of transferable simulation. 
Simulating a larger number of states mitigates this issue, though a general solution is still needed. 
Finally, applying this methodology to transferable simulation across chemical space will require extending the transferable ansatz to simultaneously accommodate systems with variable electron numbers.

While the limitations offer promising directions for future research, we have shown that transferable deep-learning QMC simulation of molecular PESs can already be applied to study challenging electronic structure problems as is.
The presented method improves the accuracy and substantially reduces the cost of simulating excited-state PESs using deep-learning QMC approaches. 
Our transferable ansatz provides a coherent framework not only for accessing energies but also, as shown, for computing intermolecular overlaps across geometries (or time steps). 
These overlaps can be used in the surface hopping picture to approximate non-adiabatic couplings and, in turn, determine hopping probabilities \cite{hammes-schiffer1994}.
At the same time, transferable simulation of excited-state PESs can be a source of highly accurate solutions to the Schrödinger equation to be consumed in down-stream tasks, such as parameterizing excited-state force fields.
\section{Methods}\label{sec11}

The time-independent Schrödinger equation
\begin{equation}\label{eq:SE}
            \hat H \Psi^i = E^i \Psi^i ~~~ \Psi^i \in \mathcal{H}^- \,,
\end{equation}
defines the electronic states of a molecule as the eigenstates $\Psi^i$ of the molecular Hamiltonian $\hat H$, indexed in order of increasing energy $E^i$.
Additionally, electrons are subject to the fermionic anti-symmetry constraint, that is the solutions have to belong to the anti-symmetric subspace of the Hilbert space $\mathcal{H}^-$.
To make the problem computationally tractable, in quantum chemistry,  one commonly works in the Born--Oppenheimer approximation, neglecting spin-orbit coupling and other relativistic effects.
For molecules this approximation is implemented by the clamped-nucleus Coulomb Hamiltonian, which in first quantization (and applying atomic units) takes the form
\begin{equation}\label{eq:hamil}
    \hat H = -\frac{1}{2} \sum_i \nabla_i^2 - \sum_{iI} \frac{Z_I}{|\mathbf{R}_{I}-\mathbf{r}_i|} + \sum_{i<j} \frac{1}{|\mathbf{r}_i-\mathbf{r}_j|} \, ,
\end{equation}
with $\nabla_i^2$ denoting the Laplace operator, ${\bf r}_i$ denoting the positions of the electrons while $Z_I$ and $\mathbf{R}_I$ are the nuclear charges and (fixed) nuclei positions.
Due to the decoupling of electronic and nuclear degrees of freedom, the Schr\"odinger equation holds for every molecular configuration independently, with a mere parametric dependency on the nuclei.
This defines the adiabatic Born--Oppenheimer potential energy surfaces
\begin{equation}\label{eq:BOPES}
    \hat H\Psi^i(\mathbf r | \mathbf{R}) = E^i(\mathbf{R})\Psi^i(\mathbf r | \mathbf{R})\,.
\end{equation}

\subsection{Variational optimization}
In order to approximate the eigenstates of the Schrödinger equation \eqref{eq:SE} we apply the variational Monte Carlo (VMC) method. 
VMC is a variational method that minimizes the Rayleigh quotient
\begin{equation}\label{eq:rayleigh_quotient}
    \boldtheta '= \underset{\boldtheta}{\text{argmin}} \frac{\braket{\Psi_\boldtheta|\hat H|\Psi_\boldtheta}}{\braket{\Psi_\boldtheta|\Psi_\boldtheta}}
\end{equation}
of a parameterized trial wave function $\Psi_\boldtheta$.
The goal is to find the optimal set of parameters $\boldtheta '$, which can then be used to evaluate expectation values.
The high dimensional integrals in \eqref{eq:rayleigh_quotient} are approximated by Monte Carlo integration with importance sampling.
This stochastic integration method poses very few restriction on the functional form of the wave function ansatz, enabling the inclusion of explicit electronic correlation.

Throughout the optimization, expectation values are estimated on finite batches of electron samples drawn from the probability density associated with the square of the wave function via Markov chain Monte Carlo.
At each iteration, the model parameters are updated with a flavor of stochastic gradient descent until convergence is reached.
Details on the VMC method and the peculiarities of optimizing neural network wave functions are found in earlier works \cite{foulkes2001,schatzle2023}.

There are various extensions of the VMC method to target excited states.
The basic requirement of these extensions is preventing the optimization from collapsing to the ground electronic state (global energy minimum), while ensuring convergence to the excited states (local energy minima).
In this work we employ a penalty-based excited state optimization method \cite{pathak2021, entwistle2023, wheeler2024}, where the loss function of each state is composed of the energy of the given state plus a term penalizing its overlap with all lower-lying states.
Additionally, a spin-penalty can be introduced to target states of a specific spin sector \cite{szabo2024}, leading to the loss function of
\begin{align}\label{eq:excited_state_optimization}
\mathcal{L}[\Psi_{\boldtheta_1}, ..., \Psi_{\boldtheta_{N_\text{s}}}&] =  \\\sum_i^{N_\text{s}}\Big[\braket{\hat{H}}^{i} + \beta &\braket{\hat{S}^2}^{i} +  \sum_{j}^{j<i} \alpha_{ij} |\braket{\Psi_{\boldtheta_i} | \bar\Psi_{\boldtheta_j}}|^2 \Big] \,. \nonumber
\end{align}
where $\hat{S}^2$ is the total spin operator and $\{\boldtheta_1,..,\boldtheta_{N_\text{s}}\}$ represent the parameters of the wave functions corresponding to each of the $N_\text{s}$ different states. The bar (in $\bar\Psi$) indicates that no gradient is taken with respect to the parameters of the specified wave function.
The factor $\beta$ scales the spin penalty with respect to the energy and overlap contributions and is set to a fixed value (here 10).
The pairwise overlap scaling parameters $\alpha_{ij}$ are determined via the same expression as used in our earlier work \cite{szabo2024}
\begin{align}
    \alpha_{ij} &= \\
    \tilde \alpha &\cdot \text{max}\left(| E^i_\mathbf{R} - E^j_\mathbf{R} |, \, \text{ewm}(\sqrt{\text{Var}(E_\mathbf{R}^{\text{loc},i})}), 10^{-3}\right) \, , \nonumber
\end{align}
where $\tilde \alpha > 1$ is the new free parameter shared between all pairs of states (set to $\tilde \alpha = 4$ in all experiments reported here), $\mathrm{ewm}(\cdot)$ denotes the exponentially weighted mean over the training iterations, $E_\mathbf{R}^{\mathrm{loc},i}$ is the batch of local energies in the current step, and $\mathrm{Var}(\cdot)$ denotes taking the variance.
The first argument of the maximum function ensures that the constraint required for obtaining orthogonal states \cite{wheeler2024} is fulfilled, while automatically scaling $\alpha_{ij}$ in a system-specific way.
The second argument prevents the collapse of the states in the earliest stages of the training where our mean energy estimates cannot yet give good approximations of the energy differences.

\subsection{Transferable optimization}
Typically, VMC is applied in the context of single-point calculations, that is the parameters of the trial wave function are optimized to minimize the energy expectation value of a specific molecular geometry.
With the vastly increased expressivity of neural network wave functions, however, the concept of transferable optimization has been introduced.
The idea consists of sharing a subset of the wave function parameters across multiple geometries \cite{scherbela2022} or making the ansatz explicitly dependent on the nuclear geometry \cite{gao2022}, and minimizing the energy for multiple molecular geometries simultaneously.
The transferable optimization makes use of the fact that under the Born--Oppenheimer approximation \eqref{eq:BOPES} the variational principle of quantum mechanics holds independently for every nuclear configuration of a molecule, that is
\begin{equation}\label{eq:transferable_optimization}
E_{0}(\mathbf R)=\min_{\Psi}\frac{\langle \Psi|\hat{H}|\Psi\rangle_\mathbf{R}}{\langle \Psi|\Psi\rangle_\mathbf{R} }\le\min_ \boldtheta \frac{\braket{\Psi_\boldtheta|\hat{H}|\Psi_ \boldtheta}_\mathbf{R}}{\braket{\Psi_\boldtheta|\Psi_ \boldtheta}_\mathbf{R}} \,,
\end{equation}
where $\braket{\cdots}_\mathbf{R}$ denotes the electronic expectation value at the nuclear geometry $\mathbf{R}$.
In this work we combine the optimization across multiple molecular geometries \eqref{eq:transferable_optimization} with the penalty based loss \eqref{eq:excited_state_optimization} and augment the wave function architecture to make it more suitable to the transferable setup.
This yields geometrically transferable wave functions capable of describing different electronic states of a molecule across a range of molecular geometries.

We generalize the training of the wave function model by iterating through a dataset of molecular geometries $\boldsymbol{\Re}$. 
This amounts to minimizing the average loss \eqref{eq:excited_state_optimization} over the whole dataset
\begin{equation}
    \boldtheta' = \underset{\boldtheta}{\text{argmin}}\sum_{\mathbf{R} \in \boldsymbol{\Re}} \mathcal{L}[\Psi_{\boldtheta_1}, ..., \Psi_{\boldtheta_{N_\text{s}}}](\mathbf R)\,.
\end{equation}
In order to sample electrons for the estimation of the expectation value of the loss function at each training iteration, we employ the Metropolis algorithm and keep independent Markov chains for each geometry of the dataset in memory.
To retain sufficiently large electron batches (electron batch sizes) for sufficiently accurate estimates of the expectation values, we subsample the dataset of molecules and alternate between successive batches (molecule batch size).
While we find the training to be fairly insensitive to the exact choice of molecule batch size, a round robin approach with a molecule batch size of one \cite{scherbela2022} can lead to the Markov chains of other molecules going stale, especially for large datasets.
Therefore, when considering larger molecule datasets, a compromise between electron batch size and molecule batch size is made (details can be found in Supplementary Section \ref{sec:app:hyperparameters}).
To account for the changes of the electron distribution due to updates of the wave function parameters, we employ a decorrelation sampler that equilibrates the Markov chains before drawing each batch of electron samples.
While the approach of keeping a stack of independent Markov chains for all molecules of the dataset convinces through its simplicity, scaling to datasets with thousands of molecular geometries it eventually hits its limitations.
In these cases the sampling can be replaced by either updating nuclear configurations within the Markov chains and warping electrons alongside the nuclei \cite{gao2023b} or initializing fresh Markov chains at every new molecular geometry \cite{foster2025}.

\subsection{Dynamic state ordering and level crossings}
\label{sec:methods:reordering}
The penalty based excited state optimization implements a ladder of states approach, where collapse to the ground state is prevented by orthogonalization of excited states with respect to lower-lying states \eqref{eq:excited_state_optimization}. 
The resulting order of states is of no relevance in single-point simulation and the approach guarantees the optimization of pure states, as opposed to the mixed states resulting from a symmetric loss function that penalizes both ansatzes.
However, when extending the loss to a transferable setting the order of the states gains importance, as the energy ordering might change at conical intersections and level crossings (right panel of Fig.~\ref{fig:ethylene}).
Albeit not strictly necessary, a continuous representation of the wave function along the geometries of the dataset is favorable both for model optimization in practice, and ease of interpreting the simulation results.
To address this, we propose a more flexible penalty method where the direction of orthogonalization adapts based on relative energies at each configuration.
We generalize the overlap penalty term of our loss function to be dependent on the nuclear coordinates as
\begin{align}\label{eq:penalty_energyordering}
    \mathcal{L}[\Psi_{\boldtheta_1}&,..., \Psi_{\boldtheta_{N_\text{s}}}](\bold{R}) = \nonumber\\
    &\,... + \sum_{j}^{E_\mathbf{R}^j< E_\mathbf{R}^i} \alpha_{ij} |\langle \Psi_{\boldtheta_i} | \bar \Psi_{\boldtheta_j}\rangle_\mathbf{R} |^2  
\\
\text{where}& ~~ E_\mathbf{R}^j < E_\mathbf{R}^i \Longrightarrow_{\mathrm{approx.}} \nonumber \\&\braket{\Psi_{\boldtheta_{j}}|\hat{H}|\Psi_{\boldtheta_{j}}}_\bold{R} < \braket{\Psi_{\boldtheta_i}|\hat{H}|\Psi_{\boldtheta_i}}_\bold{R}\nonumber\,,
\end{align}
where for simplicity we omit the sums on the configurations $\bold R$ and on the number of states $i$. Note that the re-indexing is applied independently for every molecular configuration $ \bold R \in \boldsymbol{\Re}$.
In practice we estimate the energy expectation values of \eqref{eq:penalty_energyordering} by maintaining exponential walking averages of the training energies of each state at every geometry of the dataset.
If, however, the ordering of nearly-degenerate states changes during the training, the automatic scaling of $\alpha_{ij}$ ensures that no large, discontinuous changes are made to the loss function.
By grouping states based on their physical character, this allows the network to optimize the wave functions more effectively.

While this loss does not strictly enforce the correct ordering, we find that the continuity of the wave function is typically favored due to the smoothness of the neural network ansatz and enhanced optimization dynamics, as demonstrated on Fig.~\ref{fig:ethylene}.
The continuity of the wave functions at level crossings can be probed by evaluating intermolecular overlaps
\begin{equation}\label{eq:intermolecular_overlap}
    O^{i,j}_{\mathbf{R}_m,\mathbf{R}_n}=\braket{\Psi_{\theta_i}(\mathbf{r}, \mathbf{R}_m)|\Psi_{\theta_j}(\mathbf{r}, \mathbf{R}_n)}\;,
\end{equation}
between the transferable wave function of state $i$ and $j$ at nuclear configurations $\mathbf{R}_m$ and $\mathbf{R}_n$. Fig.~\ref{fig:ethylene}c illustrates how dynamic state ordering allows the transferable network to maintain a consistent character throughout the crossing. The overlap of the excited state wave function at $\phi=90^\circ$ with itself at varying $\phi$ remains close to one, while its overlap with the ground state stays near zero. In contrast, without dynamic ordering, the network switches character at the crossing, causing the excited-state overlap to drop to zero at $\phi=100^\circ$ while the overlap with the ground state rises to one.
This effect also shows in the non-adiabatic couplings between the ground state and the excited state.

\subsection{Transferable ansatz}\label{sec:ansatz}
All reported experiments were carried out using an ansatz derived from the Psiformer architecture \cite{glehn2023} that was extended to explicitly account for changes in the molecular geometry.
The wave function $\Psi: \mathbb R^{3N}\rightarrow\mathbb R$ is parametrized via the usual many-body Slater-Jastrow form
\begin{align}
    \Psi&(\mathbf{r}|\mathbf{R}) =\\
    &\exp(J(\mathbf{r})) \sum_{k=1}^{N_\text{det}} \det\Big[\phi^k_j(\boldsymbol{h}^{\text{elec}}_i(\mathbf{r}|\mathbf{R}))\,
    \varphi^k_j(\mathbf{r}_i|\mathbf{R})\Big]_{i,j=1}^{N_\text{elec}}\,\nonumber ,
\end{align}
where $\mathbf{r} \in \mathbb R^{3N_\text{elec}}$ are the positions of the $N_\text{elec}$ electrons, $\mathbf{R}\in \mathbb R^{3N_\text{nuc}}$ stands for the coordinates of the $N_\text{nuc}$ nuclei, $J$ is a permutation variant Jastrow factor, $\phi^k_j$ are ``generalized" orbitals predicted from the many-body embedding of the $i$'th electron ${\bf h}^\text{elec}_i$, and $\varphi^k_j(\mathbf r_i|\mathbf R)$ are single-electron envelope functions anchored at the nuclei. 
The determinant of the $N_\text{elec}\times N_\text{elec}$ orbital matrix enforces fermionic antisymmetry, and the $N_\text{det}$ such determinants are summed to increase the expressivity of the ansatz.
Following the standard convention in variational Monte Carlo calculations for molecular systems, we employ spin-assigned wave functions in which the first $N^\uparrow$ electrons are assigned spin-up and the remaining $N_\text{elec}^\downarrow = N_\text{elec} - N_\text{elec}^\uparrow$ electrons are assigned spin-down. Consequently, the fixed spin configuration does not appear as an explicit argument of the wave functions but is incorporated implicitly through the spin assignment.

At the core of the wave function is the self-attention neural network architecture, which encodes the electron configuration in a latent space.
To that end, initial per-electron and per-nuclei feature vectors are constructed from rescaled electron-nuclei and nuclei-nuclei distances, electron spins and nuclei charges, respectively
\begin{align}
     & \boldsymbol{h}^{\text{elec},0}_i = \\
     &~~~~\text{MLP}^{\text{elec}}_{\text{embed}}\Big(\bigoplus_{I=1}^{N_\text{nuc}}\Big(s(\mathbf{r}_i-\mathbf{R}_I),\ s({|\mathbf{r}_i-\mathbf{R}_I|)}\Big),\sigma_i,\Big)\nonumber\\   
     & \boldsymbol{h}^{\text{nuc},0}_I = \text{MLP}^\text{nuc}_{\text{embed}}\Bigg(\sum_{I'} \text{MLP}^{\text{edge}}\Big(\\
     &~~~~~~~~\big(s(\mathbf{R}_I-\mathbf{R}_{I'}) \oplus\ s(|\mathbf{R}_I-\mathbf{R}_{I'}|)\oplus Z_I\big)\Big)\Bigg)\, ,   \nonumber
\end{align}
where $\oplus$ is the concatenation operation, $\sigma_i \in \{+1,-1\}$ is the electron spin, $Z_I \in \mathbb Z$ are the nuclear charges and MLP denotes multi-layer perceptrons.
The function $s(\cdot)$ corresponds to a logarithmic rescaling of the distances and differences
\begin{equation}
    s(\mathbf{R}) = \frac{\log(1+|\mathbf{R}|)}{|\mathbf{R}|}\mathbf{R}\,,
\end{equation}
increasing the numerical stability of the model at large interparticle distances.
Each layer then updates the embeddings via multi-headed (causal) self-attention with residual connections
\begin{align}
&\mathbf{f}_i^{\ell+1} = \mathbf{h}_i^\ell + W_o^\ell \bigoplus_h\Bigg[\frac{1}{\sqrt{d}}\sum_j
\operatorname{softmax}_j\big(\\
&~~~~~~~~~~~~~~~~~~(\mathbf{q}_i^{\ell h})^T\mathbf{k}_1^{\ell h},\dots,(\mathbf{q}_i^{\ell h})^T\mathbf{k}_{N_\text{elec}}^{\ell h}\big)\mathbf{v}_j^{\ell h}\Bigg]\nonumber\\
&\mathbf{h}_i^{\ell+1} = \mathbf{f}_i^{\ell+1} + \tanh\big(W_u^{\ell+1} \mathbf{f}_i^{\ell+1} + b^{\ell+1}\big)\,,
\end{align}
with keys, values and queries given by
\begin{equation}
\mathbf{q}_j^{\ell h}=W_q^{\ell h}\mathbf{h}_j^\ell,~~
\mathbf{k}_j^{\ell h}=W_k^{\ell h}\mathbf{h}_j^\ell,~~
\mathbf{v}_j^{\ell h}=W_v^{\ell h}\mathbf{h}_j^\ell
\end{equation}
and $\operatorname{softmax}_j(\cdot)$ being the softmax function
\begin{equation}
    \operatorname{softmax}_j(x_1,...,x_n)=\frac{\exp(x_j)}{\sum_i\exp(x_i)}\,.
\end{equation}
The weight matrices $W^\ell_{(\cdot)}$ hold the trainable parameters of the $\ell$'th layer of the encoder and $d$ is the dimension of the key, query and value vectors. 
In multi-headed self-attention several independent heads, indexed by $h$, are evaluated and concatenated.
Inspired by the Born–Oppenheimer approximation, we improve the efficiency of the Laplacian computation for the electronic degrees of freedom by restricting nuclear message passing to nuclear–nuclear interactions, while allowing nuclei to provide their states to electrons.
After several layer of interactions the final many-body embeddings of the electrons are projected to obtain the generalized orbitals $\phi^k_j = W_\text{orb} \mathbf{h}_i^\text{elec}$.
At the same time, from the final nucleus representations the envelope exponent parameters $\zeta_{Ijkn}$ are predicted, by passing them through a gated linear unit \cite{shazeer2020} with sigmoid activation.

To remedy the memory bottleneck induced by the computation of the orbital envelope functions $\varphi^k_j$ \cite{pfau2024,gao2024}, we employ envelopes similar to the simplified envelopes of Gao \cite{gao2024}, that is we compute the envelopes as
\begin{equation}
    \varphi^k_j({\bf r}_i) = \sum_{I}^{N_{\text{nuc}}} \sum_{n}^3 \exp(-\zeta_{Ijkn}|\mathbf{r}_i - \mathbf{R}_I|) \, .
\end{equation}

Furthermore, because self-attention does not directly access pairwise electron distances, it is not well suited to capture the electron–electron cusp condition \cite{kato1957}. We therefore augment it with a simple two-parameter Jastrow factor that enforces the cusp condition analytically
\begin{equation}
\begin{aligned}
    J(\mathbf{r}) =& \sum_{i<j;\,\sigma_i=\sigma_j} -\frac{1}{4}\frac{\alpha_{par}^2}{\alpha_{par}+|\mathbf{r}_i-\mathbf{r}_j|} \\&+ \; \sum_{i,j;\,\sigma_i\neq\sigma_j} -\frac{1}{2}\frac{\alpha_{anti}^2}{\alpha_{anti}+|\mathbf{r}_i-\mathbf{r}_j|}\,.
\end{aligned}
\end{equation}
The cusp parameters are trainable and initialized with $\alpha_{par}=\frac14$ and $\alpha_{anti}=\frac12$.

\subsection{Sharing parameters between electronic states}\label{sec:methods:parameter_sharing}
Parameter sharing has been demonstrated to be very efficient when transferably simulating ground states by exploiting similarities in the wave functions of different molecular configurations.  
Analogous similarities may arise when simulating multiple electronic states of a molecule, for instance due to the presence of core orbitals, which are not involved in any of the electronic excitations.
While previous works employed distinct sets of model parameters for each excited state respectively \cite{entwistle2023,szabo2024,lu2023}, we propose to share a large fraction of the model parameters between the states
\begin{equation}
\begin{aligned}
    \Psi_{\boldtheta_1}(\bold{R}),..., &\Psi_{\boldtheta_{N_\text{s}}}(\bold{R}) \\&\rightarrow \Psi_{(\boldtheta_1, \boldtheta^\text{s})}(\bold{R}),..., \Psi_{(\boldtheta_{N_\text{s}},\boldtheta^\text{s})}(\bold{R}) \,.
\end{aligned}
\end{equation}
The parameter sharing is implemented by averaging a subset of the model parameters after each gradient update.
Supposing that there is appreciable similarity between the wave functions of different states, parameter sharing improves error cancellation on relative energies by correlating the latent space representation of electrons in the wave functions of different states and effectively increases the number of gradient steps the shared parameters receive.
It is important to note that parameter sharing across electronic states is considered a useful tool for reducing simulation costs. However, it is neither necessary to attain high accuracy nor guaranteed to improve convergence, in a manner similar to related concepts such as state averaging in CASSCF or frozen-core orbitals in EOM-CC. 
Furthermore, the optimal parameter sharing strategies may depend on the electronic structure of the system under investigation and could, in principle, vary across the simulated geometries.
We empirically determine that such dependence is negligible.
Based on our experiences we recommend sharing the entire self-attention mechanism.
Occasionally, sharing large fractions of the parameters and thereby reducing the per-state expressivity of the ansatzes can hinder convergence to the correct set of states.
To account for these outliers parameter sharing can be turned off at any time throughout the optimization or the effect can be remedied by reducing the fraction of shared model parameters, for instance sharing only the initial layers of the self-attention block.

Since parameter sharing and ECPs leverage related similarities between electronic states, the benefits of sharing parameters may be reduced when comparing simulations with ECPs to their (generally more expensive) all-electron counterpart.
At the same time we did not observe a deterioration of the accuracy in relative energies when combining parameter sharing with ECPs.
An extended analysis of the parameter sharing across electronic states for the ethylene pyramidalization can be found in Supplementary Section \ref{sec:app:ablation}.

\subsection{Pseudopotentials and pseudohamiltonians}
While core electrons contribute dominantly to the total energy, they are chemically inert with most chemical and bonding properties governed by valence electrons. By removing the core electrons and representing their effect through pseudopotentials, the number of active electrons is reduced, leading to faster convergence and improved stability. 
This simplification also smoothens the valence wave function, enhancing the efficiency of Monte Carlo sampling by allowing larger step sizes, and the reduced electron count reduces the memory requirements and speeds up convergence of the wave function. 

A typical pseudopotential, such as ccECP~\cite{bennett2017} employed for the methylenimmonium cation calculations, replaces the nuclear Coulomb potential (second term in Equation \eqref{eq:hamil}) with a non-local operator acting on the angular momentum components of the wavefunction, see~\cite{li2022,schatzle2023} for details.
Although this substitution greatly simplifies the physical problem, it involves computational overhead due to the additional integrals introduced by the non-local operators, which we approximate using a 12-point quadrature scheme. 
Recently, local forms of pseudopotentials~\cite{ichibha2023} have been introduced in the context of deep-learning QMC~\cite{fu2026}, and are used here to model the rubredoxin active site. 
Such local pseudopotentials, also termed pseudohamiltonians, circumvent the costly non-local integrals while maintaining a comparable level of accuracy. 
This represents a further substantial improvement in computational efficiency, thereby enabling the scaling of deep-learning QMC calculations to systems of considerable size. 

\backmatter

\section*{Declarations}
\bmhead{Data availability}
All data used to generate the figures in this manuscript are available at https://doi.org/10.6084/m9.figshare.32935382. For convenience, molecular geometries and energies are additionally provided in the Supplementary Information.

\bmhead{Code availability}
The code used to run the DeepQMC simulations is publicly available on GitHub at https://github.com/deepqmc/deepqmc, released under the MIT license. The version used in this study corresponds to v1.3.0, archived at https://doi.org/10.5281/zenodo.21456036.

\bmhead{Acknowledgements}
Funding is gratefully acknowledged from Deutsche Forschungsgemeinschaft (DFG, German Research Foundation) under Germany’s Excellence Strategy -- The Berlin Mathematics Research Center MATH+ (EXC-2046/1, project ID: 390685689) projects (AA1-6, AA2-8, AA2-20) and Deutsche Forschungsgemeinschaft (DFG, German Research Foundation) project NO825/3-2. A. Cuzzocrea acknowledges support from the Alexander von Humboldt Foundation. The authors thankfully acknowledge the computing time granted by the Resource Allocation Board and provided on the supercomputer Lise at NHR@ZIB as part of the NHR infrastructure. The calculations presented here were conducted with computing resources under the project bem00084.

\bmhead{Author contributions}
ZS, PBS, AC and FN conceived the project.
ZS, PBS and AC developed the method in full detail.
ZS and PBS wrote the computer code with contributions from AC and MM.
The numerical experiments were conceived, carried out, and analyzed by ZS, PBS, and AC.
ZS, PBS, AC and MM wrote the manuscript with input from FN.
FN supervised the project.
Funding was acquired by AC and FN.

\bmhead{Competing interests}
The authors declare no competing interests.

\putbib[sn-bibliography]
\end{bibunit}
\newpage
\onecolumn
\appendix
\begin{bibunit}
\captionsetup[table]{skip=12pt}

\section{Numerical results}
\subsection{Ethylene}
\FloatBarrier
Numerical results of the single-point optimizations can be found in the Supplementary information of Reference \cite{szabo2024}.

\begin{table*}[h!]
\footnotesize{
    \centering
\begin{tabular}{c|cc||c|cc}
 Torsion & \multirow{2}{*}{$A^1_g$} & \multirow{2}{*}{$B^1_1u$} & Pyramidalization & \multirow{2}{*}{$A^1_g$} & \multirow{2}{*}{$B^1_1u$} \\
    angle &&& angle &&  \\
\hline\hline
0 & -78.58312(14) & -78.31030(19) & 0 & -78.38210(16) & -78.47060(14) \\
15 & -78.57895(16) & -78.31041(17) & 20 & -78.38481(14) & -78.46971(14) \\
30 & -78.56641(13) & -78.32932(15) & 40 & -78.39163(14) & -78.46422(13) \\
45 & -78.54586(14) & -78.34926(17) & 60 & -78.39816(13) & -78.45055(13) \\
60 & -78.51807(13) & -78.36443(17) & 70 & -78.39932(15) & -78.43907(14) \\
70 & -78.49716(13) & -78.37132(15) & 80 & -78.39787(14) & -78.42355(12) \\
80 & -78.47735(14) & -78.37566(15) & 90 & -78.39280(13) & -78.40462(14) \\
85 & -78.47065(16) & -78.37693(14) & 95 & -78.38907(13) & -78.39346(12) \\
90 & -78.46759(14) & -78.37736(16) & 97.5 & -78.38699(14) & -78.38793(15) \\
- & - & - & 100 & -78.38425(13) & -78.38193(14) \\
- & - & - & 102.5 & -78.38152(13) & -78.37577(15) \\
- & - & - & 105 & -78.37848(14) & -78.36934(13) \\
- & - & - & 110 & -78.37141(12) & -78.35636(16) \\
- & - & - & 120 & -78.35455(13) & -78.32961(14) \\
\end{tabular}
    \caption{Total energies in atomic units for the two lowest singlet states of the ethylene pyramidalization predicted via transferable DeepQMC optimization including dynamic state reordering, and sharing of the electron-nucleus transformer parameters.}
    \label{tab:ethylen}}
\end{table*}

\begin{table*}[h!]
\footnotesize{
    \centering
\begin{tabular}{c||rrr||rrr}
 & \multicolumn{3}{c||}{Torsion} &  \multicolumn{3}{c}{Pyramidalization}\\
Atom & x&y&z &  x&y&z \\
\hline\hline
C &-0.6750&0.0&0.0&-0.6885&0.0&0.0\\
C &0.6750&0.0&0.0&0.6885&0.0&0.0\\
H &-1.2429&0.0&-0.9303&-1.3072+cos$\,\phi\cdot$0.6187 &-sin$\,\phi\cdot$0.6187&0.9155 \\
H &-1.2429&0.0&0.9303&-1.3072+cos$\,\phi\cdot$0.6187 &-sin$\,\phi\cdot$0.6187&-0.9155 \\
H &1.2429&-sin$\,\tau\cdot0.9303$&-cos$\tau\cdot0.9303$&1.3072 &0.9155&0.0 \\
H &1.2429&sin$\,\tau\cdot0.9303$&cos$\,\tau\cdot0.9303$&1.3072 &-0.9155&0.0\\
\end{tabular}
    \caption{Molecular geometries of ethylene in \AA{}ngstr\"{o}m. $\tau$ denotes the torsion angle between the CH\textsubscript{2} groups, while $\phi$ denotes the pyramidalization angle of one of the CH\textsubscript{2} groups.}
    \label{tab:ethylen_geom}}
\end{table*}
\FloatBarrier
\subsection{Carbon dimer}
\FloatBarrier
Numerical results of the single-point optimizations can be found in the Supplementary information of Reference \cite{szabo2024}.

\begin{table*}[h!]
\footnotesize{
    \centering
    \begin{tabular}{l|cccc}
    Bond & \multirow{2}{*}{$X^1\Sigma_g^+$} & \multirow{2}{*}{$A^1\Pi_u$-} & \multirow{2}{*}{$A^1\Pi_u$+} & \multirow{2}{*}{$B^1\Delta_g$} \\
    length \\
    \hline\hline
    0.9952 & -75.76794(23) & -75.65244(24) & -75.65224(24) & -75.58232(25) \\
    1.1196 & -75.88988(22) & -75.81400(22) & -75.81446(22) & -75.75909(22) \\
    1.1818 & -75.91182(22) & -75.85150(23) & -75.85200(26) & -75.80737(21) \\
    1.2440 & -75.91711(21) & -75.87022(24) & -75.87156(22) & -75.83669(22) \\
    1.3062 & -75.91222(21) & -75.87663(22) & -75.87841(22) & -75.85244(24) \\
    1.3684 & -75.89981(23) & -75.87388(21) & -75.87556(22) & -75.85847(20) \\
    1.4928 & -75.86341(21) & -75.85390(22) & -75.85559(22) & -75.85110(24) \\
    1.6172 & -75.82856(24) &           --- & -75.82561(23) & -75.83087(23) \\
    1.7416 & -75.80365(23) &           --- & -75.79373(23) & -75.80644(23) \\
    1.8660 & -75.77953(25) &           --- & -75.76414(27) & -75.78143(24) \\
\end{tabular}
       \caption{Total energies in atomic units of the singlet states along the carbon dimer dissociation curve, predicted via transferable DeepQMC optimization including dynamic state reordering, and sharing of the electron-nucleus transformer parameters. Bond lengths are given in \AA{}ngstr\"{o}m.}
    \label{tab:carbon_dimer_singlet}}
\end{table*}

\begin{table*}[h!]
\footnotesize{
    \centering
\begin{tabular}{l|cccc}
    Bond & \multirow{2}{*}{$a^3\Pi_u-$} & \multirow{2}{*}{$a^3\Pi_u+$} & \multirow{2}{*}{$c^3\Sigma_u^+$} & \multirow{2}{*}{$b^3\Sigma_g^-$} \\
    length \\
    \hline\hline
    0.9952 & -75.69047(27) & -75.69006(26) & -75.74981(26) & -75.60561(28) \\
    1.1196 & -75.85204(23) & -75.85143(23) & -75.85860(22) & -75.79266(25) \\
    1.1818 & -75.88871(22) & -75.88807(25) & -75.87342(21) & -75.84176(22) \\
    1.2440 & -75.90741(25) & -75.90621(23) & -75.87180(24) & -75.86781(24) \\
    1.3062 & -75.91256(24) & -75.91231(23) & -75.86016(23) & -75.88178(26) \\
    1.3684 & -75.91034(23) & -75.90991(25) & -75.84132(27) & -75.88673(22) \\
    1.4928 & -75.88783(22) & -75.88824(22) & -75.79541(25) & -75.87674(21) \\
    1.6172 & -75.85593(23) & -75.85491(23) & -75.75698(26) & -75.85308(23) \\
    1.7416 & -75.82175(25) & -75.82081(24) & -75.73483(26) & -75.82495(25) \\
    1.8660 & -75.78949(25) & -75.78791(26) & -75.72195(26) & -75.79604(27) \\
\end{tabular}
       \caption{Total energies in atomic units of the triplet states along the carbon dimer dissociation curve, predicted via transferable DeepQMC optimization including dynamic state reordering, and sharing of the electron-nucleus transformer parameters. Bond lengths are given in \AA{}ngstr\"{o}m.}
    \label{tab:carbon_dimer_triplet_quintet}}
\end{table*}

\FloatBarrier
\subsection{Methylenimmonium cation}\label{sec:app:methylenimmonium}
\FloatBarrier
\begin{figure}[h!]
    \centering
    \includegraphics[width=0.65\textwidth]{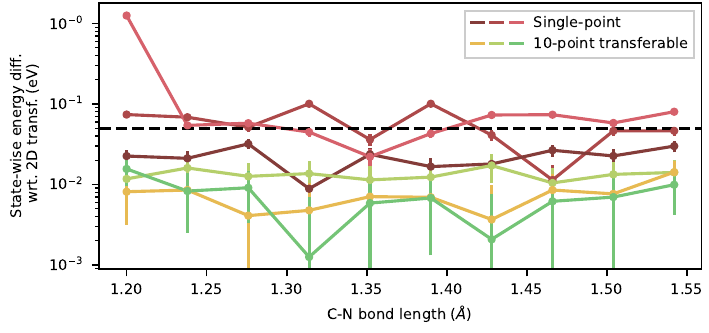}
    \caption{\textbf{Energy differences among DeepQMC models at evaluation.} 
We plot the energy difference per state for the single-point and 10-point transferable models relative to the 2D 100-point transferable model at $\tau=0$. The data used are the evaluated energies from Table~\ref{tab:ch2nh2p_single_point} (single-point) and Table~\ref{tab:ch2nh2p_10point_transferable} (10-point transferable), with the 2D transferable model at $\tau=0$ from Table~\ref{tab:ch2nh2p_100point_transferable} serving as the reference. The horizontal dashed line marks the threshold for chemical accuracy. }
    \label{fig:ch2nh2p_comparing_deepqmc}
\end{figure}
We report the numerical results for the single point, 10  and 100 point transferable optimizations of Methylenimmonium cation on the 10 geometries obtained by varying the C-N bond at $\tau=0$ in Tables \ref{tab:ch2nh2p_single_point}--\ref{tab:ch2nh2p_100point_transferable}.

On Fig.~\ref{fig:ch2nh2p_comparing_deepqmc}, we take the 100-point transferable model as a reference and plot the energy differences of the single-point and 10-point models relative to it. Notably, the single-point calculation shows a 1 eV deviation for the second excited state at \ch{C-N} = 1.2 \AA{}, indicating non-convergence. In contrast, all other states, as well as all states at other geometries, show close agreement, with the 10-point transferable model consistently matching the reference (100-point model) to better than chemical accuracy (0.05 eV), confirming the robustness of the transferable approach.

\begin{table*}[h!]
\footnotesize{
    \centering
\begin{tabular}{c|ccc}
    \ch{C-N} bond & \multirow{2}{*}{$1A_1$} & \multirow{2}{*}{$1A_2$} & \multirow{2}{*}{$2A_1$} \\
    length \\
    \hline\hline
    1.200 & -17.71035(6)  & -17.39825(8)  & -17.27392(8)  \\
    1.238 & -17.71685(6)  & -17.40894(8)  & -17.35095(9)  \\
    1.276 & -17.71901(6)  & -17.41417(8)  & -17.37426(9)  \\
    1.314 & -17.71625(7)  & -17.41772(8)  & -17.39224(8)  \\
    1.352 & -17.71198(6)  & -17.41396(8)  & -17.40434(10) \\
    1.390 & -17.70481(6)  & -17.41242(7)  & -17.41811(8)  \\
    1.428 & -17.69605(7)  & -17.40418(10) & -17.42798(8)  \\
    1.466 & -17.68604(7)  & -17.39547(10) & -17.43475(8)  \\
    1.504 & -17.67476(6)  & -17.38802(11) & -17.43846(8)  \\
    1.542 & -17.66313(7)  & -17.37816(8)  & -17.44214(8)  \\
\end{tabular}
       \caption{Total energies in atomic units of singlet states of methylenimmonium cation at $\tau=0$ and varying C-N bond length. 
        We compute standard single point calculation with the Psiformer ansatz. Bond lengths are given in \AA{}ngstr\"{o}m.
        }
    \label{tab:ch2nh2p_single_point}}
\end{table*}

\begin{table*}[h!]
\footnotesize{
    \centering
\begin{tabular}{c|ccc}
    \ch{C-N} bond & \multirow{2}{*}{$1A_1$} & \multirow{2}{*}{$1A_2$} & \multirow{2}{*}{$2A_1$} \\
    length \\
    \hline\hline
    1.200 & -17.71007(7)   & -17.39685(10)  & -17.32185(9)  \\
    1.238 & -17.71652(7)   & -17.40775(10)  & -17.35028(9)  \\
    1.276 & -17.71838(7)   & -17.41345(10)  & -17.37348(9)  \\
    1.314 & -17.71647(7)   & -17.41522(10)  & -17.39165(9)  \\
    1.352 & -17.71173(7)   & -17.41391(10)  & -17.40638(8)  \\
    1.390 & -17.70478(9)   & -17.40999(10)  & -17.41780(8)  \\
    1.428 & -17.69579(14)  & -17.40382(10)  & -17.42639(8)  \\
    1.466 & -17.68571(7)   & -17.39639(10)  & -17.43283(8)  \\
    1.504 & -17.67443(7)   & -17.38762(10)  & -17.43760(9)  \\
    1.542 & -17.66254(7)   & -17.37710(10)  & -17.44057(9)  \\
\end{tabular}
       \caption{Total energies in atomic units of singlet states of methylenimmonium cation at $\tau=0$ and varying C-N bond length. The energies are predicted via transferable optimization at the ten configurations at $\tau=0$, including dynamic state reordering, without sharing of the electron-nucleus transformer parameters. Bond lengths are given in \AA{}ngstr\"{o}m. }
   \label{tab:ch2nh2p_10point_transferable}}
\end{table*}

\begin{table*}[h!]
\footnotesize{
    \centering
\begin{tabular}{c|rrr}
    \ch{C-N} bond & \multirow{2}{*}{$1A_1$} & \multirow{2}{*}{$1A_2$} & \multirow{2}{*}{$2A_1$} \\
    length \\
    \hline\hline
    1.200 &  -17.71021(15)   & -17.39825(8)  &  -17.32098(21)  \\
    1.238 &  -17.71681(17)   & -17.40894(8)  &  -17.34968(19)  \\
    1.276 &  -17.71855(15)   & -17.41417(8)  &  -17.37286(19)  \\
    1.314 &  -17.71668(18)   & -17.41772(8)  &  -17.39132(19)  \\
    1.352 &  -17.71186(16)   & -17.41396(8)  &  -17.40588(19)  \\
    1.390 &  -17.70494(15)   & -17.41242(7)  &  -17.41726(18)  \\
    1.428 &  -17.69613(20)   & -17.40418(10) &  -17.42602(18)  \\
    1.466 &  -17.68580(15)   & -17.39547(10) &  -17.43277(18)  \\
    1.504 &  -17.67462(17)   & -17.38802(11) &  -17.43705(24)  \\
    1.542 &  -17.66277(17)   & -17.37816(8)  &  -17.43992(19)  \\
\end{tabular}
       \caption{Total energies in atomic units of singlet states of methylenimmonium cation at $\tau=0$ and varying C-N bond length. The energies are predicted via transferable optimization on the 100 point grid and evaluated on the ten configurations at $\tau=0$, including dynamic state reordering, without sharing of the electron-nucleus transformer parameters. Bond lengths are given in \AA{}ngstr\"{o}m.}  
    \label{tab:ch2nh2p_100point_transferable}}
\end{table*}

\begin{table*}[h!]
\footnotesize{
    \centering
\begin{tabular}{c||rrr}
Atom & x&y&z \\
\hline\hline
C & 0.0& 0.0 & 0.0\\
N & d & 0.0 & 0.0\\
H &-0.5445& 0.0 & 0.9431\\
H &-0.5445& 0.0 & -0.9431\\
H & d + 0.5297 &-sin$\,\tau\cdot0.9431$&-cos$\tau\cdot0.9431$\\
H & d + 0.5297 &sin$\,\tau\cdot0.9431$&cos$\tau\cdot0.9431$\\
\end{tabular}
    \caption{Molecular geometries of methylenimmonium cation in \AA{}ngstr\"{o}m. $d$ denotes the length of the C--N bond, while $\tau$ denotes the torsion angle between the CH\textsubscript{2} and NH\textsubscript{2} groups.}
    \label{tab:ch2nh2_geom}}
\end{table*}
\FloatBarrier
\subsection{Rubredoxin}\label{sec:app:rubredoxin}
\FloatBarrier
\begin{figure}[h!]
    \centering
    \includegraphics[width=0.65\textwidth]{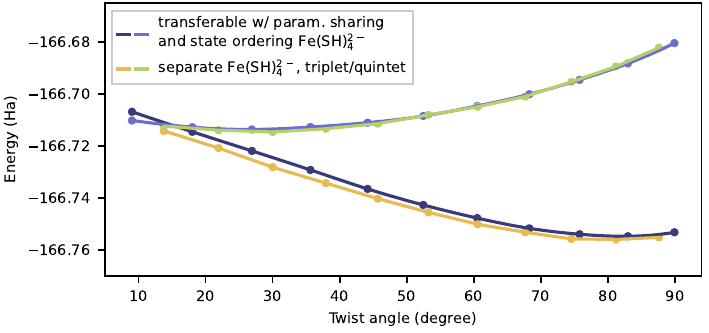}
    \caption{\textbf{Comparison of joint transferable simulation of the rubredoxin active site model with independent transferable simulation in spin sectors.} The total energies of Fe(SH)$_4^{2-}$ simulations employing a joint optimization scheme with overlap penalty and parameter sharing across electronic states, and calculating each state independently in its respective spin sector utilizing the spin penalty, are shown.}
    \label{fig:rubredoxin_alternative}
\end{figure}

To demonstrate all components of the methodology the results on rubredoxin in Fig.~5 have been obtained in a joint simulation of the quintet and triplet state, that is using the overlap penalty with dynamic state ordering, employing parameter sharing between the two electronic states and employing a spin penalty to the triplet state (the quintet state exhibits degeneracies). 
However, since the targeted states are the lowest energy states in two different spin sectors, an alternative strategy is to run two separate computations, a ground state calculation (with $S_z=2$) to obtain the quintet state and a spin-penalized calculation (with $S_z=1$) to simulate the triplet state (more details in \cite{szabo2024}). 
In Fig.~\ref{fig:rubredoxin_alternative} joint simulation is compared with independent runs in the two spin sectors.
While both results generally agree well, it can be seen that the ground state simulation for the separate runs has lower energy than its counterpart from the joint simulation. 
This is a common consequence of the error cancellation due to parameter sharing, which typically slightly increases the absolute energy of the lower-lying states at the benefit of improved relative energies.\\

\begin{figure}[h]
    \centering
    \includegraphics[width=0.65 \linewidth]{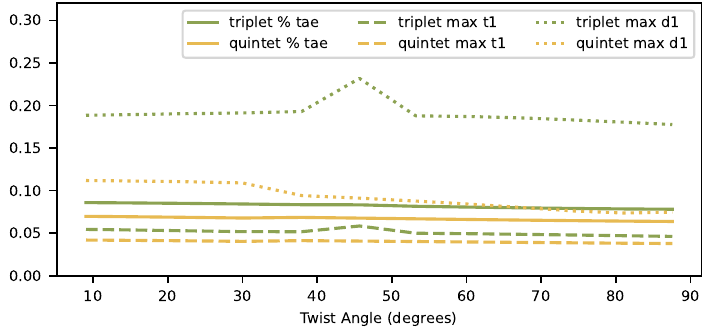}
    \caption{\textbf{Multi-reference diagnostics of the CCSD(T) simulation of the rubredoxin active site model.} The figure depicts the T1 (dashed lines) and D1 (dotted lines) diagnostics and the percentage contribution of perturbative triples to the total atomization energy (TAE) (solid lines) CCSD(T) reference simulations presented in Fig.5 of the manuscript.}
    \label{fig:rubredoxin_mr_diagonistics}
\end{figure}
Furthermore, we conducted an in-depth validation of the simulations of the Fe(SH)$_4^{2-}$ rubredoxin active site model.
Firstly, in order to assess the qualitative correctness of the single-reference coupled cluster calculation, we examined the multi-reference statistics of the CCSD(T) solution. 
We evaluated the T1 and D1 diagnostics, as well as the percentage contribution of perturbative triples to the total atomization energy (TAE), as shown in Fig.~\ref{fig:rubredoxin_mr_diagonistics}. 
The maximum T1 amplitudes per geometry are approximately 0.06 for the triplet state and 0.04 for the quintet state, while the maximum D1 amplitudes per geometry range from 0.18 to 0.23 for the triplet state and from 0.075 to 0.11 for the quintet state. 
The TAE value is approximately 0.85 for the triplet state and 0.7 for the quintet state across all geometries. 
These results suggest a moderate multi-reference character in the triplet state and a less pronounced multi-reference character in the quintet state.
As expected, the multi-reference character increases slightly towards more symmetric, planar geometries, which also exhibit the alleged intersystem crossing. 
Although the triplet diagnostics are somewhat elevated, overall the multi-reference statistics remain within the regime where CCSD(T) is expected to provide reliable qualitative results.

\begin{figure}[h]
    \centering
    \includegraphics[width=\linewidth]{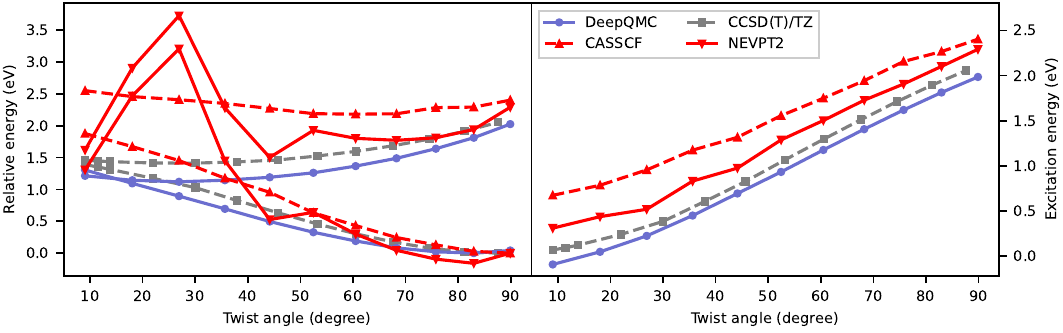}
    \caption{\textbf{CASSCF and NEVPT2 simulations of the rubredoxin active site model as qualitative validation against the CCSD(T) reference and transferable deep-learning QMC results.}  Relative energies across the PES (left panel) and excitation energies (right panel) are shown.}
    \label{fig:rubredoxin_mrref}
\end{figure}
To provide additional confirmation of the correct qualitative behavior, we simulated the PES using the multi-reference CASSCF method, followed by NEVPT2 treatment to model dynamic electron correlations.
The simulations were run with PySCF\cite{sun2020}, with the def2-TZVP Karlsruhe basis set for S, C, and H, and the Stuttgart RSC-1997 ECP and valence basis set for the Fe atom.
To obtain suitable active spaces we have employed AVAS\cite{sayfutyarova2017} on the restricted open-shell Hartree-Fock solution, selecting the Fe 3d orbitals and the S 3p orbitals.
This results in an active space comprising 28 electrons in 17 orbitals.
In order to balance the accuracy of the triplet and quintet states, both states were treated in a joint simulation employing state averaging with a triplet and a quintet solver, respectively.
The CASSCF computation was then followed by two independent NEVPT2 simulations.
One of the major challenges in applying CASSCF to excited state PESs is maintaining a consistent description of the electronic structure across molecules and spin states. 
In order to align the orbital representation of the electronic structure and obtain a consistent active space across the geometries, we began our simulations from a central geometry on the PES, sequentially simulating the neighboring geometries by projecting the previous solutions onto the new atomic orbitals as an initial guess.

Fig.~\ref{fig:rubredoxin_mrref} shows the simulations of the Fe(SH)$_4^{2-}$ rubredoxin active site model obtained with CASSCF and NEVPT2.
The CASSCF and NEVPT2 excitation energies are in qualitative agreement with the CCSD(T) and DeepQMC simulations and indicate the validity of the results.
CASSCF captures the overall trends observed in CCSD(T) and DeepQMC, removing the intersystem crossing between triplet and quintet state predicted by DFT. 
With the inclusion of NEVPT2 corrections, differences in electronic structure representations (e.g., orbitals and active spaces) lead to deviations in absolute energies between the molecular geometries. 
However, when expressed as excitation energies, NEVPT2 shows substantially improved agreement with CCSD(T) and DeepQMC. 
Moreover, in regions where the electronic structure is represented more consistently, closer agreement between the CCSD(T) and DeepQMC PESs is observed.
Inconsistent representations of the electronic structure across the PES are a well-known issue within a CASSCF-NEVPT2 workflow, that we were unable to fully address.
Overcoming these difficulties is one of the key aspects that we intend to solve using our transferable DeepQMC approach.

\begin{table*}[h!]
\footnotesize{
    \centering
\begin{tabular}{c|cccc}
    \multirow{2}{*}{Twist angle} &  \multicolumn{2}{c}{Quintet} & \multicolumn{2}{c}{Triplet} \\
     &$\braket{\hat{H}}$ & $\braket{\hat{S}^2}$  & $\braket{\hat{H}}$ & $\braket{\hat{S}^2}$ \\
    \hline\hline
\,\,\,
9.05  & -166.7068(2) & 5.99859(18) & -166.7102(2) & 2.00135(7) \\
18.04 & -166.7145(2) & 5.99851(17) & -166.7127(2) & 2.00125(6) \\
26.92 & -166.7219(2) & 5.99840(17) & -166.7136(2) & 2.00109(6) \\
35.64 & -166.7292(2) & 5.99844(16) & -166.7127(2) & 2.00109(6) \\
44.16 & -166.7365(2) & 5.99824(16) & -166.7110(2) & 2.00113(6) \\
52.46 & -166.7427(2) & 5.99836(16) & -166.7084(2) & 2.00121(6) \\
60.50 & -166.7478(2) & 5.99821(16) & -166.7046(2) & 2.00108(6) \\
68.26 & -166.7517(2) & 5.99827(15) & -166.7001(2) & 2.00111(6) \\
75.75 & -166.7540(2) & 5.99831(15) & -166.6945(2) & 2.00119(6) \\
82.96 & -166.7548(2) & 5.99840(16) & -166.6882(2) & 2.00136(6) \\
89.89 & -166.7533(2) & 5.99818(17) & -166.6803(2) & 2.00121(7) \\
\end{tabular}
   \label{tab:rubredoxin_joint}
    \caption{Total energies and spin expectation value in atomic units of simulations of the rubredoxin active site model for varying twist angles. The energies are obtained via transferable optimization with the overlap penalty and parameter sharing.}}
\end{table*}

\begin{table*}[h!]
\footnotesize{
    \centering
\begin{tabular}{c|cc}
    Twist angle &  Quintet & Triplet \\
    \hline\hline
\,\,\
13.79 & -166.7142(2) & -166.7123(2) \\
21.96 & -166.7208(2) & -166.7139(2) \\
30.02 & -166.7282(3) & -166.7146(2) \\
37.94 & -166.7343(2) & -166.7133(2) \\
45.68 & -166.7403(2) & -166.7114(2) \\
53.23 & -166.7456(2) & -166.7080(2) \\
60.56 & -166.7502(2) & -166.7049(2) \\
67.66 & -166.7533(2) & -166.7009(2) \\
74.53 & -166.7558(3) & -166.6953(2) \\
81.17 & -166.7560(2) & -166.6894(2) \\
87.58 & -166.7552(2) & -166.6821(2) \\
\end{tabular}
        \caption{Total energies in atomic units of simulations of the rubredoxin active site model for varying twist angles. The energies are obtained from two independent simulations in the quintet and triplet sector, respectively.}  
\label{tab:rubredoxin_independent}
}
\end{table*}

\begin{table*}[h!]
\footnotesize{
    \centering
\begin{tabular}{c||lll}
Atom & \multicolumn{3}{c}{Geometry (z-matrix)} \\
\hline\hline
Fe & - & - & - \\
S  &0 $\text{d}_{\text{S-Fe}}$ & - & - \\
S  &0 $\text{d}_{\text{S-Fe}}$& 1 $\alpha_\text{S-Fe-S}$& -\\
S  &0 $\text{d}_{\text{S-Fe}}$& 1 $\alpha_\text{S-Fe-S}$& 2 $\beta_\text{S-Fe-S-S}$\\
S  &0 $\text{d}_{\text{S-Fe}}$& 2 $\alpha_\text{S-Fe-S}$& 1 $\beta_\text{S-Fe-S-S}$\\
H  &1 1.33&0 $\alpha_\text{Fe-S-H}$&4 $\beta_\text{S-Fe-S-H}$\\
H  &2 1.33&0 $\alpha_\text{Fe-S-H}$&3 -$\beta_\text{S-Fe-S-H}$\\
H  &3 1.33&0 $\alpha_\text{Fe-S-H}$&2 -$\beta_\text{S-Fe-S-H}$\\
H  &4 1.33&0 $\alpha_\text{Fe-S-H}$&1 $\beta_\text{S-Fe-S-H}$\\

\end{tabular}
    \caption{Molecular geometries of the rubredoxin active site model in \AA{}ngstr\"{o}m. The values of $\text{d}_\text{S-Fe}\in[2.2634, 2.3534]$, $\alpha_\text{Fe-S-H}\in[2.0227,1.7890]$ and $\beta_\text{S-Fe-S-H}\in[1.4884, 1.2127]$ are obtained by linear interpolation (joint, 11 steps) between the reference minima for the triplet and quintet states, respectively. The reference quintet and triplet minimum geometries are obtained from the work of dePolo et al. \cite{depolo2016}. $\alpha_\text{S-Fe-S}=\text{arccos}\big(-\text{cos}(\gamma)^2\big)$ and $\beta_\text{S-Fe-S-S}=\text{arccos}\big(-\text{sin}(\gamma)^2/(1+\text{cos}(\gamma)^2)\big)$ are obtained from linear interpolation (joint, 11 steps) of $\gamma\in[1.5149, 0.9562]$. The ``twist angle'' is defined as the angle between the two planes S$^1$-Fe-S$^2$ and S$^3$-Fe-S$^4$.}
    \label{tab:rubredoxin_geom}}
\end{table*}

\FloatBarrier
\section {Parameter-sharing analysis}\label{sec:app:ablation}
\FloatBarrier
\begin{figure}[h!]
    \centering
    \includegraphics[width=\linewidth]{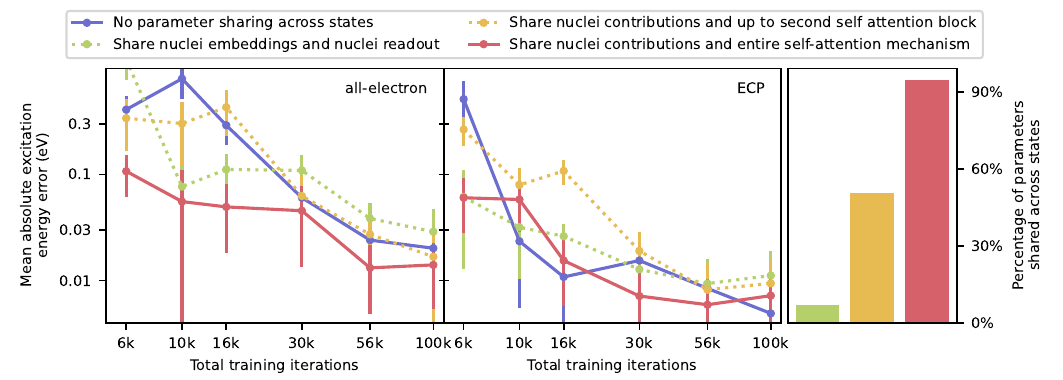}
    \caption{\textbf{Parameter sharing study on ethylene pyramidalization.} The mean absolute error in excitation energy with respect to the reference for the 14 geometries of the ethylene pyramidalization PES, for all-electron simulations (left panel) and with effective core potentials (middle panel) is shown. Each data point in the plot consists of 14 evaluation runs for a transferable wave function at various checkpoints. Different colors represent different parameter sharing schemes with increasingly many parameters being shared from independent ansatzes (blue) to the sharing of the entire transformer (red). The rightmost panel shows the ratio of parameters being shared in each configuration.}
    \label{fig:ablation}
\end{figure}
To provide more insights on the effect of parameter sharing across electronic states we investigate the impact of different parameter sharing-strategies on the convergence of the excitation energies along the ethylene pyramidalization and assess the interaction of ECPs and parameter sharing.
First, on the left panel of Fig. \ref{fig:ablation} we validate that parameter sharing across electronic states enhances error cancellation and accelerates optimization.
We observe that the advantageous effect of parameter sharing diminishes with an increasing number of training steps.
This asymptotic convergence is expected since in the limit of approaching the exact eigenstates, all simulations yield equivalent errors.
Then, on the middle panel of Fig. \ref{fig:ablation} we repeat the experiment with core electrons replaced by ECPs, demonstrating that ECPs can be combined with parameter sharing without a loss in accuracy, but that the efficiency gains may be smaller.
Overall we compare four different parameter sharing modes, with the ratio of shared parameters shown on the right panel of Fig. \ref{fig:ablation}: no parameter sharing across states, sharing only nuclear embeddings and the nuclear readout  ($\sim$7\% of parameters), sharing nuclear embeddings, nuclear readout and the first two self-attention layers ($\sim$51\% of parameters) and sharing the entire self-attention mechanism ($\sim$95\% of parameters).
All of the runs for this experiment were performed transferably across the fourteen molecular geometries of the ethylene pyramidalization PES, respectively.
To assess the convergence of the excitation energies along the PES we evaluate checkpoints of the transferable wave function at various stages of the training and compare the excitation energies to the reference energies of Pfau et al. \cite{pfau2024}.

On average a trend of equal or enhanced convergence with an increased number of shared parameters is observed, albeit the stochasticity of the optimization prevents a clear point-wise ordering of the excitation energy error based on the fraction of parameters being shared. 
For the simulation of the two lowest singlet states of ethylene we find that sharing all of the core neural networks and having states differ only in the final projection to the orbitals gives the best performance in relative energies, but it is conceivable that larger systems or simulations with more excited states benefit more from a different choice.

\section{Training hyperparameters}\label{sec:app:hyperparameters}
A summary of the most important training hyperparameters used for the presented single-point and transferable DeepQMC calculations can be found in Table \ref{tab:hyperparams}.
Other hyperparameters were set identically to the corresponding single-point calculations, as described by Szab\'o \textit{et al.} \cite{szabo2024}.

\begin{table*}[h!]
\small{
    \centering
    \begin{tabular}{cc|cccc}
        System                        & variant      & \makecell{pretrain\\steps} & \makecell{total train\\steps} & \multicolumn{2}{c}{batch size} \\
                                      &              &                &                   & geometry & electron            \\ \hline \hline
        \multirow{2}{*}{ethylene}     & single-point &             1k &              450k / 700k &        1 &                2048 \\
                                      & transferable &              0 &              200k &        4 &                 512 \\ \hline
        \multirow{2}{*}{carbon dimer} & single-point &             1k &              500k &        1 &                2048 \\
                                      & transferable &              0 &              200k &        1 &                2048 \\ \hline             
        \multirow{3}{*}{\ch{CH2NH2+}} & single-point &              0 &                1M &        1 &                 2048 \\
                                      & transferable 10 conf.&              0 &               50k &       10 &                1024 \\ 
                                      & transferable 100 conf.&              0 &               100k &       20 &                 512 \\ \hline
        \multirow{2}{*}{\ch{Fe(SH)^{2-}_4}}               & transferable joint &              0 &              200k &                                      1 &                4096 \\& transferable separate &              0 &              400k &                                      1 &                2048 \\\hline
    \end{tabular}
    \caption{Training hyperparameters used in this work. All other transferable hyperparameters are identical to the respective single-point calculations, further details on which can be found in Reference \cite{szabo2024}. In case of single-point iterations, the number of total training steps refers to the sum of the training steps performed for each of the geometries of the given dataset (9 geometries for ethylene torsion, 14 for pyramidalization, 10 for carbon dimer dissociation, and 10 for the methylenimmonium cation).  }
    \label{tab:hyperparams}}
\end{table*}

\section{Computational setup}
All presented simulations are run with the \textsc{DeepQMC} software package \cite{schatzle2023,deepqmc}, which provides \textsc{jax} \cite{bradbury2018} implementations of the core algorithms with GPU acceleration and multi-node parallelization.
Parameter updates are computed using the KFAC algorithm \cite{martens2015,botev2022}.
For the efficient computation of the kinetic energy operator $\sum_i\nabla^2_i$ we use automatic differentiation and employ the forward Laplacian framework \cite{li2024a, gao2023}.
All experiments were run on NVIDIA-A100 graphics cards on the supercomputer Lise at NHR.
\twocolumn
\putbib[sn-bibliography]

\end{bibunit}

\end{document}